\documentstyle[12pt]{article}
\begin{document}
\def\baselinestretch{1.4}
\setlength{\oddsidemargin}{0pt}
\setlength{\evensidemargin}{0pt}
\setlength{\marginparwidth}{0pt}
\setlength{\marginparsep}{10pt}
\setlength{\topmargin}{0pt}
\setlength{\headheight}{0pt}
\setlength{\headsep}{0pt}
\setlength{\footheight}{12pt}
\setlength{\footskip}{30pt}
\setlength{\textheight}{630pt}
\setlength{\textwidth}{470pt}
\setlength{\columnsep}{10pt}
\setlength{\columnseprule}{0pt}
\begin{flushright}
UT-637
\\
April 1993
\end{flushright}
\begin{center}
{\large\bf Instanton and Spectral Flow   \\
             in \\
Topological Conformal Field Theories} \vspace{.7 in}\\
{\bf Toshio Nakatsu
            \footnote{Address after 20th April:
                      Research Institute for Mathematical Sciences,
                      Kyoto university, Kyoto 606, Japan.}
            \footnote{A Fellow of the Japan Society for the Promotion
                      for Japanese Junior Scientists.
                      Partly supported by the Grant-in-Aid for
                      Scientific Research from the Ministry of Education,
                      Science and Culture (No. 04-2597).}
         and
    Yuji Sugawara
            \footnote{A Fellow of the Japan Society for the Promotion
                      for Japanese Junior Scientists.
                      Partly supported by the Grant-in-Aid for
                      Scientific Research from the Ministry of Education,
                      Science and Culture (No. 05-3531).}
                                                       \vspace{.5 in}\\
        {\it Department of Physics, University of Tokyo}\\
        {\it Bunkyo-ku,Tokyo 113,Japan \vspace{.8 in}}}

\end{center}

\newcommand{\br}{{\bf R}}
\newcommand{\bc}{{\bf C}}
\newcommand{\bz}{{\bf Z}}
\newcommand{\bq}{{\bf Q}}
\newcommand{\bn}{{\bf N}}
\newcommand{\Om}{\Omega}
\newcommand{\df}{\stackrel{\rm def}{=}}
\newcommand{\co}{{\scriptstyle \circ}}
\newcommand{\de}{\delta}
\newcommand{\si}{\sigma}
\newcommand{\ep}{\varepsilon}
\newcommand{\gerg}{\mbox{g}}
\newcommand{\gert}{\mbox{t}}
\newcommand{\germ}{\mbox{m}}
\newcommand{\dimn}{\mbox{dim}}
\newcommand{\om}{\omega}
\newcommand{\al}{\alpha}
\newcommand{\la}{\lambda}
\newcommand{\La}{\Lambda}
\newcommand{\De}{\Delta}
\newcommand{\vphi}{\varphi}
\newcommand{\Exp}{\mbox{Exp}}
\newcommand{\gerh}{\mbox{h}}
\newcommand{\gerb}{\mbox{b}}
\newcommand{\gera}{\mbox{a}}
\newcommand{\gerA}{{\cal A}}
\newcommand{\rank}{\mbox{rank}}
\newcommand{\bm}[1]{\mbox{\boldmath ${#1}$}}
\newcommand{\ca}[1]{{\cal #1}}
\newcommand{\ger}[1]{\mbox{#1}}
\newcommand{\lb}{\lbrack}
\newcommand{\rb}{\rbrack}
\newcommand{\rn}[1]{\romannumeral #1}
\newcommand{\msc}[1]{\mbox{\scriptsize #1}}
\newcommand{\dsp}{\displaystyle}
\newcommand{\scs}[1]{{\scriptstyle #1}}
\newcommand{\deebar}{\bar{\partial}}

\newcommand{\sdprod}
  {\hbox{$\hspace{1mm}\rule{0.2mm}{2mm}\hspace{-0.7mm} \times$}}
\newcommand{\nprod}{\hspace{-1mm}~^{\scs{\circ}}_{\scs{\circ}}}

\newcommand{\larr}{\leftarrow}
\newcommand{\rarr}{\rightarrow}
\newcommand{\Larr}{\Leftarrow}
\newcommand{\Rarr}{\Rightarrow}
\newcommand{\llarr}{\longleftarrow}
\newcommand{\lrarr}{\longrightarrow}
\newcommand{\Llarr}{\Longleftarrow}
\newcommand{\Lrarr}{\Longrightarrow}
\newcommand{\lerarr}{\leftrightarrow}
\newcommand{\Lerarr}{\Leftrightarrow}
\newcommand{\llerarr}{\longleftrightarrow}
\newcommand{\Llerarr}{\Longleftrightarrow}

\newcommand{\bra}[1]{\langle #1 |}
\newcommand{\ket}[1]{| #1 \rangle}

\newcommand{\dplus}{\hbox{$\hspace{1mm} + \hspace{-3.4mm}
   \stackrel{\co}{\rule{0cm}{1.5mm}} \hspace{1.5mm} $}}
\newcommand{\dminus}{\hbox{$\hspace{1.3mm} - \hspace{-3.5mm}
    \stackrel{\co}{\rule{0cm}{1mm}}\hspace{1.5mm} $}}

\newcommand{\scdplus}{\hbox{${\scriptstyle\hspace{0.5mm} + \hspace{-1.7mm}
   \stackrel{\co}{\rule{0cm}{1mm}} \hspace{0.5mm}} $}}
\newcommand{\scdminus}{\hbox{${\scriptstyle\hspace{0.5mm} - \hspace{-1.9mm}
    \stackrel{\co}{\rule{0cm}{0.8mm}}\hspace{0.7mm}}$}}

\newcommand{\be}{\begin{equation}}\newcommand{\ee}{\end{equation}}
\newcommand{\bea}{\begin{eqnarray}} \newcommand{\eea}{\end{eqnarray}}
\newcommand{\ba}[1]{\begin{array}{#1}} \newcommand{\ea}{\end{array}}

\newcommand {\eqn}[1]{(\ref{#1})}
\newcommand {\eq}[1]{eq.~(\ref{#1})}
\newcommand {\eqs}[1]{eqs.~(\ref{#1})}
\newcommand {\Eq}[1]{Eq.~(\ref{#1})}
\newcommand {\eql}[1]{eqs.~(\ref{#1},}
\newcommand {\eqr}[1]{\ref{#1})}
\newcommand {\eqm}[1]{\ref{#1},}

\renewcommand{\normalbaselines}{\baselineskip20pt \lieskip3pt
\lineskiplimit3pt}

\newcommand{\mapru}[1]{\smash{\mathop{\hbox to 1cm{\rightarrowfill}}
\limits^{#1}}}
\newcommand{\maprd}[1]{\smash{\mathop{\hbox to 1cm{\rightarrowfill}}
\limits_{#1}}}
\newcommand{\maplu}[1]{\smash{\mathop{\hbox to 1cm{\leftarrowfill}}
 \limits^{#1}}}
\newcommand{\mapld}[1]{\smash{\mathop{\hbox to 1cm{\leftarrowfill}}
 \limits_{#1}}}
\newcommand{\maprud}[2]{\smash{\mathop{\hbox to 1cm{\rightarrowfill}}
 \limits^{#1}_{#2}}}
\newcommand{\maplud}[2]{\smash{\mathop{\hbox to 1cm{\leftarrowfill}}
 \limits^{#1}_{#2}}}

\newcommand{\lmapru}[1]{\smash{\mathop{\hbox to 1.5cm{\rightarrowfill}}
\limits^{#1}}}
\newcommand{\lmaprd}[1]{\smash{\mathop{\hbox to 1.5cm{\rightarrowfill}}
\limits_{#1}}}
\newcommand{\lmaplu}[1]{\smash{\mathop{\hbox to 1.5cm{\leftarrowfill}}
 \limits^{#1}}}
\newcommand{\lmapld}[1]{\smash{\mathop{\hbox to 1.5cm{\leftarrowfill}}
 \limits_{#1}}}
\newcommand{\lmaprud}[2]{\smash{\mathop{\hbox to 1.5cm{\rightarrowfill}}
 \limits^{#1}_{#2}}}
\newcommand{\lmaplud}[2]{\smash{\mathop{\hbox to 1.5cm{\leftarrowfill}}
 \limits^{#1}_{#2}}}

\newcommand{\mapdl}[1]{\Big\downarrow
 \llap{$\vcenter{\hbox{$\scriptstyle#1\,$}}$ }}
\newcommand{\mapdr}[1]{\Big\downarrow
 \rlap{$\vcenter{\hbox{$\scriptstyle#1$}}$ }}
\newcommand{\mapul}[1]{\Big\uparrow
 \llap{$\vcenter{\hbox{$\scriptstyle#1\,$}}$ }}
\newcommand{\mapur}[1]{\Big\uparrow
 \rlap{$\vcenter{\hbox{$\scriptstyle#1$}}$ }}
\newcommand{\mapdlr}[2]{\Big\downarrow
 \llap{$\vcenter{\hbox{$\scriptstyle#1\,$}}$ }
 \rlap{$\vcenter{\hbox{$\scriptstyle#2$}}$ }}
\newcommand{\mapulr}[2]{\Big\uparrow
 \llap{$\vcenter{\hbox{$\scriptstyle#1\,$}}$ }
 \rlap{$\vcenter{\hbox{$\scriptstyle#2$}}$ }}

\newfont{\bg}{cmr10 scaled \magstep4}
\newcommand{\bigzerol}{\smash{\hbox{\bg 0}}}
\newcommand{\bigzerou}{\smash{\lower1.7ex\hbox{\bg 0}}}

\newcommand{\bigstarl}{\smash{\hbox{\bg *}}}
\newcommand{\bigstaru}{\smash{\lower1.7ex\hbox{\bg *}}}

\newcommand{\cleqn}{\setcounter{equation}{0}}
\renewcommand{\theequation}{\thesection. \arabic{equation}}

\newtheorem{thm}{Theorem}[section]
\newtheorem{prop}[thm]{Proposition}
\newtheorem{lem}[thm]{Lemma}
\newtheorem{cor}[thm]{Corollary}

\newpage

\begin{abstract}
A class of two-dimensional topological conformal field theories
(TCFTs) is studied within the framework of gauged WZW models
in order to obtain some insights on the global geometrical nature
of TCFTs.
The BRST quantizations of topological $G/H$ gauged WZW models
( the twisted versions of supersymmetric gauged WZW models )
are given under fixed back-ground gauge fields.
The BRST-cohomology of the system is investigated, and then
the correlation functions among these physical observables
are considered under the instanton back-grounds.
As a consequence, two-dimensional $BF$ gauge theoretical
aspects of TCFTs are revealed.
Especially, it is shown that two correlation functions under
the different instanton back-grounds can change to each other.
This process of transmutation is described by the spectral flow.
The flow is formulated as a "singular" gauge transformation
which creates an appropriate back-ground charge on the physical
vacuum of the system.
The field identification problem of the system is also discussed
from the above point of view.
\end{abstract}


\newpage

\section{Introduction}

{}~

            Two-dimensional topological conformal field theories
(TCFT) play an important role in our recent understanding of
string theory and two-dimensional gravity \cite{witten,DVV}.
An algebraic method to construct a model of TCFTs was given in
\cite{EY}; the "twist" of a model of $N=2$ superconformal
field theories (SCFTs).
It is known that a large class of $N=2$ SCFTs is obtainable
by the  constructions of Kazama and Suzuki \cite{KaS}, that is,
the supercoset constructions associated with compact K\"{a}hler
homogeneous spaces $G/H$.
The corresponding TCFTs were studied in \cite{EY2} from the
algebraic point of view.

              It is also known that supercoset CFTs
can be realized by
"supersymmetric gauged WZW models" \cite{witten,Nakatsu,Schnitzer}.
Much have been learned about coset models by realizing them
as gauged WZW models.
Especially the global geometry of coset models,
which is hidden in the free field realizations
\cite{DF},
appears  \cite{gw}.
Thus the twisted versions of the supersymmetric gauged WZW models,
that is, the topological gauged WZW models give us a chance to
study the TCFTs from the geometrical point of view.
In spite of this utility the topological gauged WZW models have been
little examined.
The purpose of this paper is to study these topological models
from the global geometrical point.

            We begin in section 2 by formulating the topological
gauged WZW model for a general compact K\"{a}hler homogeneous space
$G/H$.
The path-integral quantization of the model is given under a
fixed back-ground gauge field,
by utilizing the techniques developed in \cite{gk,ks,ns}.
Subsequently, based on the gauge-fixed form,
the local operator formulation is studied and the algebraic structure
of the model is shown.
In section 3 we investigate the spectrum of the physical observables,
that is, the BRST-cohomology of the gauge-fixed system.
The semi-classical analysis of the system provides a starting point.
Then, for example,  it arises the question whether this semi-classical
approximation is valid in our model.
Is the correlation function among these semi-classical observables
meaningful under any back-ground gauge field ?
This question is clarified in the next section.
For this purpose we turn our eyes to the BRST-cohomology
on the full Hilbert space of the system.
We introduce spectral flows $\ca{U}$
as the symmetry transformations of our model.

      In section 4 we study the correlation functions among
the physical observables
$\{ O_{a}  \}_{a \in I}$
under the instanton back-grounds.
It is shown that two correlation functions under the different
instanton back-grounds change to each other (\ref{main}):
\bea
\langle \ca{U}_{\gamma_1}(O_{a_1}) \cdots
        \ca{U}_{\gamma_N}(O_{a_N}) \rangle_{c_1}
=
\langle O_{a_1} \cdots O_{a_N}
\rangle_{c_1+\sum_{i=1}^{N}\gamma_i}.
\nonumber
\eea
The vectors $\gamma_i$ label the spectral flows and the
$c_1$ is the set of the Chern numbers of the back-ground gauge field.
This relation tells us that the instantons transumute
the physical observables  and vice versa.
It is described by the flows $\ca{U}$.
The field identification in the algebraic CFT approach is also
discussed from this rather topological view point.
Finally section 5 is devoted to our conclusions and speculations.
In appendix A the quantizations of the supersymmetric gauged WZW models
are summarized.
The correspondence between the "twist" of the gauged WZW models and
that of $N=2$ SCFTs are explained.
The former is performed by replacing the physical Weyl spinors
of the models with unphysical ghost fields,
while the latter is done by twisting $N=2$ SCAs
by their $U(1)$-currents, not the fermion number currents.
In appendix B the basic properties of spectral flows are described.
They are arranged in order to fit
our study of the topological gauged WZW models.

{}~

\section{Topological Gauged WZW Models Associated with General Compact
K\"{a}hler Homogeneous Spaces}
\cleqn

{}~

                  We shall begin with preparing some notations of Lie algebras
needed for our discussion.

                Let $G$ be a compact simple Lie group, $H$ be a closed
subgroup of $G$,and $\gerg$, $\gerh$ be the corresponding Lie algebras.
Assume that the homogeneous  space $G/H$ is
a compact K\"{a}hler space, which implies this space is a so-called
``flag manifold''. Especially
$H$ includes the maximal torus $T$ of $G$.
The topological conformal models
we investigate  in this paper will be defined
associated with the coset space of this type, or the pair $(G,H)$.
So it may be useful  to  give here a short digression  to
survey  the  structure
of $\gerg^{\bc}$ and  $\gerh^{\bc}$.

                   First $\gerh^{\bc}$ can be decomposed as follows;
\be
\gerh^{\bc} = Z(\gerh^{\bc}) \oplus \gerh_0^{\bc}
            = Z(\gerh^{\bc}) \oplus \gerh_0^{(1)\bc} \oplus \cdots
     \oplus \gerh_0^{(r)\bc}
\label{hczh}
\ee
where $Z(\gerh^{\bc})$ is the center of $\gerh^{\bc}$
and $\gerh_0^{\bc}$ is a complex semi-simple Lie algebra.
We denote its simple factors by $\gerh_0^{(1)\bc}, \ldots , \gerh_0^{(r)\bc}$.
The corresponding decomposition of $H$ is
\be
\ba{lll}
H & =&  H_0 ~\times ~\underbrace{U(1)~\times ~\cdots~ \times U(1)}
                  _{\msc{$l$ times}}  \\
  & = & H_0^{(1)} ~ \times~ \cdots ~ \times H_0^{(r)}~\times
        U(1)~\times ~\cdots~ \times U(1)  .
\ea
\label{hu1h0}
\ee
$H_0$ is the compact semi-simple Lie group corresponding to $\gerh_0^{\bc}$,
and $l = \rank \gerg - \mbox{rank}\gerh
\equiv \mbox{dim} Z(\gerh^{\bc})$.
We denote the  root system of $\gerg^{\bc}$, $\gerh_0^{\bc}$
by $\De \equiv
\De^+ \coprod \De^-$,
$\De_{\gerh} \equiv \De_{\gerh_0}^+ \coprod \De_{\gerh_0}^-$
respectively.
We also set $\Delta_{\germ}^{\pm} =
\Delta^{\pm} \setminus \De_{\gerh}^{\pm}$.
$\gerg^{\bc}$ can be decomposed into the following pieces;
\be
\ba{lll}
\gerg^{\bc} &=& \gerh^{\bc}~ \oplus~ \germ_+~ \oplus~ \germ_- \\
            &=& Z(\gerh^{\bc})~\oplus
                   ~ \gerh_0^{\bc}~ \oplus~ \germ_+~ \oplus~ \germ_-
\label{parabolic}
\ea
\ee
where
${\dsp \germ_{\pm} = \sum _{\al \in \De_{\germ}^{\pm}} \gerg_{\al} }$.
$\gerg_{\al}$ is the root space for
$\al \in \De$.
This is an example of  parabolic decompositions of $\gerg^{\bc}$
(generalizations of the Cartan decomposition);
\be
\left\{
\ba{lll}
   \lb \gerh^{\bc},~ \germ_{\pm} \rb &\subset &\germ_{\pm} \\
   \lb \germ_{\pm},~ \germ_{\pm} \rb &\subset & \germ_{\pm}
\ea
\right. .
\label{com rel h m}
\ee
If the homogeneous space $G/H$ is hermitian symmetric,
we further obtain the additional commutation relations;
\be
  \lb \germ_{\pm},~ \germ_{\mp} \rb ~\subset ~ \gerh^{\bc},
\ee
which implies that
the subalgebras $\germ_{+}$, $\germ_{-}$ necessarily become abelian.
This case  is rather  easy to handle in many respects.

           In the following part of this paper we also assume that
the group $G$ is simply-laced.

{}~

            Now we can present the model we shall work on, the "topological
gauged WZW model for the homogeneous space $G/H$".
We may call it the "topological $G/H$ model".
It is no other than the twisted version
of the $N=2$ supersymmetric gauged WZW model
\cite{witten}, \cite{ns} corresponding to the Kazama-Suzuki
supercoset model \cite{KaS}.
Let $\Sigma$ a Riemann surface.
The partition function $Z$ of the model is given by
\bea
&& Z = \int \ca{D}g \ca{D}A
\ca{D}(\chi, \bar{\chi}, \psi, \bar{\psi} )
     \, \exp \left\lb -kS_G(g:~A) - \frac{1}{2 \pi i} \int_{\Sigma}
  \{ (\deebar_{A}\psi, \chi )-(\bar{\chi}, \partial_{A}\bar{\psi})\}
   \right\rb . \nonumber \\
&& ~~~~~~~
\label{top g/h gwzw model}
\eea
In this expression
the chiral field $g$ is $G$-valued,
the gauge field $A$ is $\gerh$-valued, and
the ghost system
is defined as follows;
\bea
\ba{l}
\mbox{ghosts}~~~~
\psi~:~ \mbox{$\germ_+$-valued (0,0)-form} ,~~~
\bar{\psi} \equiv \psi^{\dag} ~:~ \mbox{$ \germ_-$-valued (0,0)-form}, \\
\mbox{anti-ghosts}~~~~
\chi~:~ \mbox{$\germ_-$-valued (1,0)-form},~~~
\bar{\chi} \equiv \chi^{\dag}~:~ \mbox{$\germ_+$-valued (0,1)-form}.
\ea
\nonumber
\eea
$D_A = \partial_A + \deebar_A$ is the canonical splitting of
the covariant exterior derivative $D_A$
defined by the complex structure
``compatible'' with the background metric $\bm{g}$ on $\Sigma$
(i.e. such that $\bm{g}$ becomes K\"{a}hler).
The inner product ``$( ~,~)$'' is the Cartan Killing form normalized by
$(\theta, \theta) =2$ ($\theta$ is the highest root of $\gerg^{\bc}$.),
``$\dag$'' is the
``hermitian conjugation'' so that $\gerg = \{ u \in \gerg^{\bc}~;~
         u ^{\dag} = -u \}$.
$S_G (g:~ A)$ stands for the action of the $G$-WZW model
gauged by the subgroup $H$;
\be
\ba{l}
\dsp
 S_{G} (g:~A) =  \frac{ i}{4 \pi}\int_{\Sigma}
  (g^{-1}\deebar g, g^{-1}\partial g)
  -\frac{i}{24 \pi}\int_{B}(\tilde{g}^{-1}d\tilde{g},
          \lb \tilde{g}^{-1}d\tilde{g}, \tilde{g}^{-1} d\tilde{g} \rb ) \\
\dsp     ~~~~~ + \frac{i}{2\pi}
  \int_{\Sigma}\{ -(g^{-1} \deebar g , A^{10}) + (A^{01}, \partial g g^{-1})
       -(A^{01}, Ad(g) A^{10}) + (A^{01}, A^{10}) \} .
\ea
\label{gwzw action}
\ee
Besides the corresponding gauge symmetry the model
(\ref{top g/h gwzw model}) satisfies
the following BRST symmetry,
which is originated in the SUSY of the untwisted model;
\be
\ba{l}
\ba{ll}
  \delta_{G/H} \chi = k\Pi_{\germ_-}\left( \partial_A g\, g^{-1} +
  \lb \chi, \psi \rb \right ),& ~~~
   \bar{\delta}_{G/H} \bar{\chi}
     = -k \Pi_{\germ_+} ( g^{-1}\deebar_A g - \lb \bar{\chi},
    \bar{\psi}\rb ) ,\\
\delta_{G/H}g = \psi g,& ~~~
  \bar{\delta}_{G/H}g = - g \bar{\psi},\\
\dsp \delta_{G/H}\psi = \frac{1}{2} \lb \psi, \psi \rb,& ~~~
\dsp \bar{\delta}_{G/H}\bar{\psi} =
 \frac{1}{2} \lb \bar{\psi}, \bar{\psi} \rb ,
\ea \\
\mbox{other combinations are defined to vanish,}
\ea
\label{susy brst}
\ee
where $\Pi_{\germ_+}$, $\Pi_{\germ_-}$ mean the orthogonal projections
onto the spaces
$\germ_+$, $\germ_-$ with respect to the inner product $(~,~)$.
$\de_{G/H},\bar{\de}_{G/H}$ satisfy the following nilpotencies;
\bea
\de_{G/H}^2=\bar{\de}_{G/H}^2=0,~~~~
\{ \de_{G/H},~\bar{\de}_{G/H} \}=0.
\eea
Because
the ghost system $\psi$, $\bar{\psi}$,
$\chi$, $\bar{\chi}$ corresponds to that of the BRST symmetry
(\ref{susy brst}), the physical Hilbert space of the model
should be restricted by taking the BRST cohomology for this symmetry.

\subsection{Path Integration}

{}~

             Let us perform the path-integration in (\ref{top g/h gwzw model})
along the line in \cite{ns}.
For that purpose we should extract the gauge volume of the
underlying $H$-gauge symmetry from (\ref{top g/h gwzw model}).
We shall parametrize the $\gerh$-gauge field $A$ as
\be
\ba{l}
A=^{h^{-1}}\! a \\
(^{h^{-1}}\! a^{01}
        =h^{-1}a^{01}h+h^{-1}\bar{\partial}h, ~^{h^{-1}}a^{10}
        =-(~^{h^{-1}}a^{01})^{\dag}),
\ea
\ee
where $a$ is a background $\gerh$-gauge field and
$a^{10},a^{01}$ are the holomorphic and anti-holomorphic components
of $a$. $h$ is a $H^{\bc}$-chiral gauge transformation \cite{gk}.
Then we insert the following identity into \eqn{top g/h gwzw model};
\be
1 = \int \ca{D}a \ca{D}h^{\dag}\ca{D}h\,
  \delta(A -^{h^{-1}}a) \De_{\msc{FP}}(~^{h^{-1}}a),
\ee
where ``$\De_{\msc{FP}}(\, ^{h^{-1}}a)$''
denotes the Fadeev-Popov (FP) determinant obtained by the change of
integration variables.
Taking account of the decomposition  \eqn{hczh} or \eqn{hu1h0},
let us express $a$, $h$ as follows;
\be
a=(i\om, a_0),~~~h=(e^{\frac{1}{2}(X + iY)}, h_0 (\equiv \rho^{1/2}U)).
\label{bggauge}
\ee
$i\om$, $e^{\frac{1}{2}(X + iY)}$ are the components corresponding
to $Z(\gerh^{\bc})$ in (\ref{hczh}).
$\om$ is a $\br^l$-valued 1-form.
$X$, $Y$ are $\br^l$-valued  scalar fields,
which correspond to the axial  and vectorial components
of the $Z(H)^{\bc}$-chiral gauge transformation.
$a_0$, $h_0$ are the components corresponding to
$\gerh_0^{\bc}$  in (\ref{hczh}).
$h_0 = \rho^{1/2}U$
($\rho \equiv h_0h_0^{\dag} \in H_{0}^{\bc}/H_0$,$U \in H_0$)
is the "polar decomposition" of $H_0^{\bc}$.

                        Under the parametrization $A=\, ^{h^{-1}}a$
the model will suffer the chiral anomaly.
We must estimate it for the dynamical variable $g$
and the ghost system $\chi$, $\bar{\chi}$, $\psi$,
$\bar{\psi}$  independently.

                       For $g$, by means of the Polyakov-Wiegmann identity
(see  \cite{pw},\cite{gk,ks,ns}) we can get
\be
\ba{lll}
S_G (g:\,^{h^{-1}}a ) &=&
 S_G (\,^{h}g:~ a) -S_G (hh^{\dag} :~ a)  \\
  &=& S_G (\,^{h}g:~ a)  -  S_{H_0} (\rho :~ a_0) \\
    & &\dsp  ~~~-\frac{i}{4\pi}\int_{\Sigma} (\deebar X, \partial X)
     +\frac{1}{2\pi} \int _{\Sigma} (X, F(\om)),
\ea
\ee
where $~^{h}g=hgh^{\dag}$.
$F(\om)\equiv d \om$ is the $Z(\gerh)$-component of curvature of the
back-ground gauge field $a=(i \om, a_0) \equiv
(i\om , a_0^{(1)},\ldots , a_0^{(r)})$.
$\dsp S_{H_0}(\rho,a_0)=
\sum_{i=1}^r S_{H_0^{(i)}} (\rho^{(i)} , a_0^{(i)})$ with
$\dsp \rho \equiv (\rho^{(1)},\ldots , \rho^{(r)})
\in \prod_{i=1}^r \left( H_0^{(i)\bc}/H_0^{(i)} \right)$.

                                For the ghost system,
the same discussion as in \cite{ns} gives;
\bea
Z_{\chi \psi} &= &
    \int \ca{D}(\chi,\bar{\chi}, \psi, \bar{\psi})\,
    \exp \left\lb -\frac{1}{2\pi i} \int_{\Sigma} \{
    (\deebar_{^{h^{-1}}a}\psi, \chi )-
    (\bar{\chi}, \partial_{^{h^{-1}}a}\bar{\psi}) \}
        \right\rb  \nonumber \\
    &=& \int \ca{D}(\chi,\bar{\chi}, \psi, \bar{\psi})\,
    \exp \left\lb -\frac{1}{2\pi i} \int_{\Sigma} \{
    (\deebar_{a}\psi, \chi )-(\bar{\chi}, \partial_{a}\bar{\psi})\} \right\rb
    \nonumber \\
    & &\dsp  ~~~\times \prod_{i=1}^r
   \exp \left \{(g^{\vee}-h_i^{\vee}) S_{H_0^{(i)}}(\rho^{(i)} :~ a_0^{(i)})
       \right\} \nonumber \\
     & & ~~~\times \exp \left\lb
     \frac{i g^{\vee}}{4 \pi} \int_{\Sigma} \{
      (\deebar X,\partial X)
     -2i (X, F(\om))
     +\frac{2}{i g^{\vee}}  \rho_{\gerg /\gerh} (X) R(\bm{g}) \} \right\rb ,
    \nonumber \\
&& ~~~~~~~
\label{determinant chi psi}
\eea
where $g^{\vee}(h_i^{\vee})$ are the dual coxeter numbers of
$\gerg^{\bc} (\gerh_{0}^{(i)\bc})$, and we introduce the notation
$\rho_{\gerg/\gerh}$ by
\bea
\rho_{\gerg /\gerh} = \rho_{\gerg}-\sum_{i=1}^r \rho^{(i)}_{\gerh_0}.
\eea
$\rho_{\gerg}$ ($\rho^{(i)}_{\gerh_0}$)
are the Weyl vectors of $\gerg^{\bc}$
($\gerh^{(i)\bc}_0$).
$R(\bm{g})$ is the Riemannian curvature tensor,
$\dsp \chi(\Sigma)=\frac{1}{2 \pi}\int R(\bm{g})$.
The contribution of $\rho$ in (\ref{determinant chi psi})
will be obtained by calculating the Schwinger term of the underlying
$\gerh_0$-current algebra.
The contribution of $X$ in (\ref{determinant chi psi}) seems
to be a ``background charge''.
It will be estimated by some direct anomaly calculation or
applying the index theorem to the corresponding Dolbeault complex.

                     We must further estimate the anomaly of the
FP determinant  $\De_{\msc{FP}}(^{h^{-1}}a)$.
It can be rewritten in the local functional form by
introducing the additional FP ghosts;
$\xi$ ($\gerh^{\bc}$-valued (0,0)-form),
$\bar{\xi}$ ($\gerh^{\bc}$-valued (0,0)-form),
$\zeta$ ($\gerh^{\bc}$-valued (1,0)-form),
$\bar{\zeta}$ ($\gerh^{\bc}$-valued (0,1)-form);
\be
\De_{\msc{FP}}(^{h^{-1}}a)=
\int \ca{D}(\zeta, \bar{\zeta}, \xi, \bar{\xi})
   \,  \exp \left\lb  - \frac{1}{2 \pi i} \int_{\Sigma}
   \{ (\deebar_{^{h^{-1}}a}  \xi, \zeta)
-(\bar{\zeta},\partial_{^{h{-1}}a} \bar{\xi})   \} \right\rb.
\ee
The chiral anomaly of this ghost system
can be computed in the same way
as that of the $\chi\psi$-ghosts;
\bea
\De_{\msc{FP}}(^{h^{-1}}a)& = &
\int \ca{D}(\zeta, \bar{\zeta}, \xi, \bar{\xi})
   \, \exp \left\lb - \frac{1}{2 \pi i} \int_{\Sigma}
   \{ (\deebar_{a_0}  \xi, \zeta)
-(\bar{\zeta},\partial_{a_0} \bar{\xi})   \} \right\rb \nonumber\\
  & &\dsp  ~~~\times \prod_{i=1}^{r}
\exp \left \{ 2 h_i^{\vee} S_{H_0^{(i)}} (\rho^{(i)}:~ a_0^{(i)})  \right\}.
\label{determinant zeta xi}
\eea
Notice that this determinant does not depend on $\om$
since $Z(\gerh^{\bc})$ acts trivially on the space
$\gerh^{\bc}\equiv Z(\gerh^{\bc}) \oplus \gerh_0^{\bc} $.

                  By summing up the above estimations of the chiral anomaly
and then dropping the gauge volumes
$\dsp \int \ca{D}U$, $\dsp \int \ca{D}Y$ off
we can obtain the following gauge fixed form of the model
(\ref{top g/h gwzw model});
\bea
 Z_{\msc{g.f.}} & = & \int \ca{D}a Z_{\msc{g.f.}}
                                                  \lb a \rb \equiv
     \int \ca{D}(a_0, \om) \,
       Z_{\msc{g.f.}}\lb a_0, \om \rb,  \nonumber \\
Z_{\msc{g.f.}} \lb a_0, \om \rb &=&
\int \ca{D}(g, \rho, X, \chi,\bar{\chi},
    \psi,\bar{\psi}, \zeta, \bar{\zeta}, \xi, \bar{\xi} )  \nonumber \\
  &  &  ~~\times \exp \left \{
       -k S_G ( g:~ a)  -
     S_{\chi \psi} (\chi,\bar{\chi}, \psi,\bar{\psi}:~ a )
      \right\} \nonumber \\
&  & \dsp  ~~\times \exp \left\{
   \sum_{i=1}^r (k + g^{\vee}+ h_i^{\vee}) S_{H_0^{(i)}}
          (\rho^{(i)}:~ a_0^{(i)})
             \right.  \nonumber  \\
& & \dsp  \left.   ~~~~~~~~~~~~- S_X (X :~ \om )
      - S_{\zeta \xi} (\zeta, \bar{\zeta}, \xi, \bar{\xi} :~ a_0) \right\}.
                \nonumber \\
&&~~~~~~~~~~~~~~~ \label{gauge fixed}
\eea
In \eqn{gauge fixed} we introduce the following notations;
\bea
&& S_{\chi \psi} (\chi,\bar{\chi}, \psi,\bar{\psi}: a )
 =  \frac{1}{2\pi i} \int_{\Sigma} \{
    (\deebar_{a}\psi, \chi )-(\bar{\chi}, \partial_{a}\bar{\psi})\}
\label{action chi psi}\\
&& S_{\zeta \xi} (\zeta, \bar{\zeta}, \xi, \bar{\xi} : a_0)
 =  \frac{1}{2 \pi i} \int_{\Sigma} \{ (\deebar _{a_0} \xi, \zeta)
-(\bar{\zeta},\partial_{a_0} \bar{\xi})   \}
\label{action zeta xi}\\
&& S_X (X : \om )
 =  \frac{ 1}{4 \pi i} \int_{\Sigma} \left\{ (\deebar X,\partial X)
     +   2i \al_+ (X, F(\om))
     + 2i\al_-  \rho_{\gerg /\gerh} (X) R(\bm{g}) \right\}, \nonumber \\
   & & ~~~~~~~~~~~~~~  \label{action X}\\
   & & ~~~~~~~(\al_+ = \sqrt{k+ g^{\vee}},~
\al_-= -\frac{1}{\sqrt{k+g^{\vee}}})  , \nonumber
\eea
where we have rescaled the scalar field $X$ as
$\al_+ X~ \rarr~ X$ in \eqn{action X}.

          Except for the ``Chern number dependent term''
$\dsp \sim \int_{\Sigma} \, (X,F(\om))$ in \eqn{gauge fixed},
The similar expression for the gauge fixed
form \eqn{gauge fixed} is given in \cite{AGSY}.
But this ``Chern number dependent term''
possesses a topological information of the system,
and will play an important role in  this paper.
We will argue on this point in section 4.

\subsection{Local Operator Formulation}

{}~

                   Nextly we will give some operator formulation of
the gauge fixed system (\ref{gauge fixed}).
We shall work on some fixed local holomorphic coordinate patch
$(U,z) \subset \Sigma$,
where we set the background fields
$\bm{g}$, $a_0$, $\om$  trivial.
The quantization of the system is straightforward,
since everything is expressed either in free fields, or
in terms of ungauged WZW models on this patch.

                              First of all, from \eqn{gauge fixed},
the total energy-momentum (EM) tensor $T_{\msc{tot}}$
of the gauge fixed system is given by ;
\be
T_{\msc{tot}} = T_g + T_{\rho} + T_X + T_{\chi \psi}+ T_{\zeta \xi},
\label{totalT}
\ee
where $T_g$, $T_{\rho}$
are the Sugawara EM tensors of the
$G_k$, $\dsp \prod_{i=1}^r \left(H_0^{\bc}/H_0\right)
    _{-(k+g^{\vee}+h_i^{\vee})}$-WZW models, and $T_X$, $T_{\chi \psi}$,
$T_{\zeta \xi}$ are those obtained from the actions
(\ref{action chi psi})-(\ref{action X}).
Their explicit forms are given as follows;
\bea
T_g &=& \frac{1}{2(k+g^{\vee})} \nprod (J_g, J_g)\nprod , \label{tg} \\
T_{\rho} & = & \sum_{i=1}^r T_{\rho}^{(i)}  ,\\
T_{\rho}^{(i)} &= &
\frac{1}{2\{ -(k+g^{\vee}+ h_i^{\vee}) + h_i^{\vee}\}}
  \nprod(J_{\rho}^{(i)}, J_{\rho}^{(i)})\nprod
= -\frac{1}{2(k+g^{\vee})} \nprod (J_{\rho}^{(i)},
                     J_{\rho}^{(i)})\nprod  \nonumber \\
&& ~~~~~~~~~~~~~~~\label{trho} \\
T_X & = & - \frac{1}{2} :(\partial_z X, \partial_z X):
    + \al_- \rho_{\gerg/\gerh}(\partial_z^2 X) \label{tx} \\
T_{\chi \psi} & = & - :(\chi_z , \partial_z \psi) :\label{t chi psi} \\
T_{\zeta \xi} & = & - :(\zeta_z , \partial_z \xi) :, \label{t zeta xi}
\eea
where
$:~~:$ denotes the standard normal ordering prescription
defined by mode expansions, and
$\dsp \nprod   A(w) B(w) \nprod = \frac{1}{2\pi i} \oint_{w} \, dz\,
\frac{A(z)B(w)}{z-w}$.
We denote the $G_k$, $H^{(i)}_{0,-(k+g^{\vee}+h_i^{\vee})}$-currents
of the corresponding WZW models by
$J_{g}=-k \partial g \, g^{-1},
J^{(i)}_{\rho}=(k+g^{\vee}+h_i^{\vee})\partial \rho^{(i)} \rho^{(i)-1}$.
One  can easily check that $T_{\msc{tot}}$
indeed has vanishing central charge;
\bea
c_{\msc{tot}} &=& c_g +c_{\rho}+ c_X + c_{\chi \psi} + c_{\zeta \xi}
\nonumber\\
  & = & \frac{k \dimn{\gerg}}{k+g^{\vee}} +
  \sum_{i=1}^{r} \frac{-(k+g^{\vee}+h_i^{\vee})
  \dimn \gerh_0^{(i)}}{-(k+g^{\vee}+h_i^{\vee})+ h_i^{\vee}}
      + l + 12 \al_-^2 \rho_{\gerg/\gerh}^2 \nonumber \\
 & & ~~~~~~+ (-2) \times \frac{1}{2}(\dimn \gerg -\dimn \gerh_0 -l)
   + (-2)\times (\dimn \gerh_0 +l) =0. \nonumber \\
&& ~~~~~~
\eea
This estimation suggests the total system
is topologically invariant.

                   Let us introduce the BRST-charges
which will characterize the physical Hilbert space
of the gauge fixed system.
The BRST symmetry
(supersymmetry) \eqn{susy brst}
should correspond to  the following BRST charge;
\be
Q_{G/H} =   \frac{1}{2\pi i} \oint dz \, G^+_{G/H}  ,
\ee
where the BRST current $G^+_{G/H}$ is defined as
\bea
G^+_{G/H} = -\al_- :(\psi, J_g + \frac{1}{2} J_{\chi \psi}):.
\label{g+gh}
\eea
We set $J_{\chi \psi}=- \mbox{[} \chi_z, \psi \mbox{]}$.
Meanwhile,
the  BRST-charges  for
the $H_0^{\bc}$ and $Z(H^{\bc})$-chiral gauge
transformations are given by
\bea
Q_{H_0^{\bc}} = \frac{1}{2\pi i} \oint dz \, G^+_{H_0^{\bc}},~~~~
Q_{Z(H^{\bc})} =
    \frac{1}{2\pi i} \oint dz \, G^+_{Z(H^{\bc})} ,
\label{qhc}
\eea
where the BRST currents $G^+_{H_0^{\bc}},
G^+_{Z(H^{\bc})}$ are defined by
\bea
G^+_{H_0^{\bc}}  & =&
   -\frac{\al_-}{\sqrt{2}}
    : (\xi, \hat{J}^{\gerh_0} + J_{\rho} +
      \frac{1}{2} J_{\zeta \xi}): , \label{g+h0c}\\
G^+_{Z(H^{\bc})} & = &
      -\frac{\al_-}{\sqrt{2}} \left \{
      (\xi, \hat{J}^{Z(\gerh)}+ J_X )
       - 2\rho_{\gerg/\gerh}(\partial_z \xi) \right \}.
\label{g+hc}
\eea
In the above expressions
we introduce the current $\hat{J}$ as
\bea
\hat{J}&=& J_g +J_{\chi \psi}
\label{hatj}
\eea
and the notations
$\hat{J}^{\gerh}(=\hat{J}^{\gerh_0}+\hat{J}^{Z(\gerh)}),
\hat{J}^{\gerh_0},\hat{J}^{Z(\gerh)}$
denote the projections of $\hat{J}$ onto the corresponding spaces.
This combined current $\hat{J}$ is $Q_{G/H}$-invariant.
The current $J_X = \al_+ \partial_z X$
is obtained from $S_X$ \eqn{action X} by taking the variation of the
$Z(\gerh)$-back-ground gauge field $i \om$.
Because the currents $\hat{J}^{\gerh_0(i)}$,
$J_{\rho}$, $J_{\zeta \xi}$, $\hat{J}^{Z(\gerh)}$  and
$J_{X}$ have levels
$k+g^{\vee}-h_i^{\vee},-(k+g^{\vee}+h_i^{\vee}),2h_i^{\vee},
k+g^{\vee}$ and $-(k+g^{\vee})$ respectively,
the "total $\gerh_0^{\bc}$, $Z(\gerh^{\bc})$-currents"
of the gauge fixed system,
\bea
J^{\gerh_0}_{\msc{tot}}  \equiv  \hat{J}^{\gerh_0} + J_{\rho} +
  J^{\gerh_0}_{\zeta \xi},~~
 J^{Z(\gerh)}_{\msc{tot}}  \equiv  \hat{J}^{Z(\gerh)} + J_X
\label{jhtot}
\eea
generate  $\gerh_0^{\bc}$, $Z(\gerh^{\bc})$-current algebras
with vanishing levels.
This implies  the nilpotency of the BRST-charges
$Q_{H_0^{\bc}}$ , $Q_{Z(H^{\bc})}$ (\ref{qhc}) (c.f. \cite{ks}).
These three BRST-charges $Q_{G/H}$, $Q_{H_0^{\bc}}$, $Q_{Z(H^{\bc})}$
anti-commute with one another.
This fact is indeed natural, since they correspond to independent
gauge degrees of freedom.

                  Set the total BRST charge of the gauge fixed system as
\bea
Q_{\msc{tot}}=
Q_{G/H}+Q_{H_0^{\bc}}+Q_{Z(H^{\bc})}.
\label{totalQ}
\eea
Then we can show that
the total EM tensor (\ref{totalT}) itself is BRST-exact,
\bea
T_{\msc{tot}}=\{ Q_{\msc{tot}},~G_{\msc{tot}}^{-} \},
\label{exacttotT}
\eea
where $G_{\msc{tot}}^{-}$ has the following factorized form:
$G^-_{\msc{tot}}= G^-_{G/H}+G^-_{H_0^{\bc}}+G^-_{Z(H^{\bc})}$.
Each element is given by ;
\bea
G^-_{G/H}  &=& -\al_- :(\chi_z , J_g + \frac{1}{2} J_{\chi \psi}): ,
          \label{G-GH} \\
G^-_{H_0^{\bc}} &=& - \frac{\al_-}{\sqrt{2}}
    (\zeta_z , \hat{J}^{\gerh_0}-J_{\rho})
         \label{G-HO} \\
G^-_{Z(H^{\bc})} & = &
      -\frac{\al_-}{\sqrt{2}} \left \{
      (\zeta_z , \hat{J}^{Z(\gerh)}-  J_X )
       - 2\rho_{\gerg/\gerh}(\partial_z \zeta_z ) \right \}.
                \label{G-ZH}
\eea
The total
$\gerh_0^{\bc},Z(\gerh^{\bc})$-currents
(\ref{jhtot}) are also BRST-exact;
\bea
J^{\gerh_0}_{\msc{tot}}=\{ Q_{\msc{tot}},
\sqrt{2}\al_+ \zeta_z^{\gerh_0} \},~~
 J^{Z(\gerh)}_{\msc{tot}}
 = \{ Q_{tot}, \sqrt{2} \al_+ \zeta_z ^{Z(\gerh)} \},
\label{exactJtot}
\eea
where $\zeta^{\gerh_0},~ \zeta^{Z(\gerh)}$ are
the $\gerh_0^{\bc},Z(\gerh^{\bc})$-components of the antighost field $\zeta$.
The BRST-exactness of $T_{\msc{tot}},
J_{\msc{tot}}^{\gerh_0,Z(\gerh)}$ will assure the topological
invariance of the system.

{}~

                   To end this section,
it is convenient to give several remarks.

                 Firstly we notice that the total EM tensor
(\ref{totalT}) can be decomposed into the following three commuting
pieces (c.f (\ref{exacttotT})-(\ref{G-ZH}));
\bea
T_{\msc{tot}}=T_{G/H}+T_{H_0^{\bc}}+T_{Z(H^{\bc})}.
\label{Ttot2}
\eea
Each element in the RHS of \eqn{Ttot2} is given by;
\bea
T_{G/H} & \equiv &
  \frac{1}{2(k+g^{\vee})} \left \{ \nprod(J_g, J_g)\nprod -
    \nprod(\hat{J}^{\gerh},\hat{J}^{\gerh} )\nprod \right \}
      + \frac{1}{k+g^{\vee}} \, \rho_{\gerg/\gerh}
         (\partial_z \hat{J}^{Z(\gerh)})
         - :(\chi_z, \partial_z \psi ):   \nonumber \\
  &=& \{  Q_{G/H},~~G_{G/H}^{-} \},   \label{tgh} \\
T_{H_0^{\bc}} & \equiv&
  \frac{1}{2(k+g^{\vee})} \left\{
    \nprod(\hat{J}^{\gerh_0}, \hat{J}^{\gerh_0})\nprod
      - \nprod (J_{\rho}, J_{\rho})\nprod  \right\}
          - :(\zeta_z^{\gerh_0}, \partial_z \xi^{\gerh_0}): \nonumber \\
   &=& \{  Q_{H_0^{\bc}},~~G_{H_0^{\bc}}  \}, \label{th0c} \\
T_{Z(H^{\bc})} & \equiv&
  \frac{1}{2(k+g^{\vee})} \nprod (\hat{J}^{Z(\gerh)},
        \hat{J}^{Z(\gerh)})\nprod
                - \frac{1}{k+g^{\vee}}\,
          \rho_{\gerg/\gerh}(\partial_z \hat{J}^{Z(\gerh)}) \nonumber \\
        & & ~~~~ - \frac{1}{2}:(\partial_z X, \partial_z X)  :
             + \al_- \rho_{\gerg/\gerh}(\partial_z^2 X)
         - :(\zeta_z^{Z(\gerh)}, \partial_z \xi^{Z(\gerh)}): \nonumber \\
   &=& \{  Q_{Z(H^{\bc})},~~G_{Z(H^{\bc})}^{-} \},   \label{tzhc}
\eea
where $\xi^{\gerh_0}$ and $\xi^{Z(\gerh)}$ denote the
$\gerh_0^{\bc},Z(\gerh^{\bc})$-components of the ghost field $\xi$.
By introducing the $U(1)$-currents;
\bea
  J_{G/H} & = & :( \psi, \chi_z): +
    \frac{2}{k+g^{\vee}} \, \rho_{\gerg/\gerh}(\hat{J}^{Z(\gerh)}),
   \label{jgh} \\
 J_{H^{\bc}_0} & = & : (\xi^{\gerh_0} , \zeta_z^{\gerh_0}):,
   \label{jh0} \\
  J_{Z(H^{\bc})} & = & : (\xi^{Z(\gerh)}, \zeta_z^{Z(\gerh)} ) :
     - \frac{2}{k+g^{\vee}} \, \rho_{\gerg/\gerh} (J_X),
   \label{jzh}
\eea
one can find out that
the pairs
$\{ G^{\pm}_{G/H}, T_{G/H}, J_{G/H} \} $,
$\{ G^{\pm}_{Z(H^{\bc})}, T_{Z(H^{\bc})}, J_{Z(H^{\bc})} \} $
generate the topological conformal algebras (TCAs)
with background charges
$\dsp Q_{\msc{KS}}=
\dimn \germ_+ -\frac{4}{k+g^{\vee}} \rho_{\gerg/\gerh}^2$,
and $\dsp Q_{\msc{CG}}=l - \frac{4}{k+g^{\vee}} \rho_{\gerg/\gerh}^2$
respectively
\footnote{A TCA $\{G^{\pm},T,J \}$ will be called a ``TCA with background
charge $Q$'' if they satisfy
$\dsp T(z)J(w)  \sim  -\frac{Q}{(z-w)^3} + \frac{1}{(z-w)^2}J(w)
+\frac{1}{z-w}\partial_w J(w)$.}.
The OPEs  of the ``residual sector''
$\{ G^{\pm}_{H_0^{\bc}}, T_{H^{\bc}_0}, J_{H^{\bc}_0} \} $
also have an almost similar form to the TCA
with back-ground charge
$Q = \dimn \gerh_0^{\bc}$.
But it is not the completely same,
since the nilpotency of the $G^-$-operator is broken;
$G^-_{H_0^{\bc}}(z)G^-_{H_0^{\bc}}(w) \not \sim 0$.
It may reflect the fact that
the manifold $H_0^{\bc}$ is not necesarily K\"{a}hler,
while $G/H$, $Z(H^{\bc})= \bc^* \times \cdots \times \bc^*$
have natural K\"{a}hler structures.

          These three algebras
commute (anti-commute for ferminonic currents) with one another.
They correspond to independent
degrees of freedom related to the different
gauge symmetries. We may call the TCA
$\{ G^{\pm}_{G/H}, T_{G/H}, J_{G/H} \} $
as that of ``Kazama-Suzuki sector'',
because the field realizations  \eqn{g+gh},
\eqn{G-GH},\eqn{tgh}, \eqn{jgh} are same as the twisted version of
the $N=2$ SCA of the Kazama-Suzuki model \cite{KaS} for $G/H$.
The TCA
$\{ G^{\pm}_{Z(H^{\bc})}, T_{Z(H^{\bc})}, J_{Z(H^{\bc})} \} $
will be called that of  Coulomb-gas(CG) sector \cite{ns}.
This is because one can write it in the following convenient form.
Introduce a $i Z(\gerh) \cong \br^l$-valued real compact boson $\vphi$
with the radius $\al_+$ (normalized by
$\dsp (u, \partial_z \vphi(z)) (v, \partial_w \vphi(w)) \sim
-\frac{(u, v)}{(z-w)^2}$, $ u, v \in Z(\gerh^{\bc})$)
and rewrite $\hat{J}^{Z(\gerh)}$ as\footnote
     {To complete the definition of $\vphi$ we should further
   define the zero-mode $\vphi_0$ appropriately.
    We take the following convention;
    $\dsp \lb a_0 , \vphi_0 \rb = - \frac{i}{2}$ with
    $\dsp a_0 =\frac{1}{2\pi i} \oint i \partial_z \vphi dz$,
     and $\lb N_{\chi \psi}, \vphi_0 \rb = 0$
    with $\dsp N_{\chi\psi}=\frac{1}{2\pi i} \oint :(\psi(z),\chi(z)):dz$.}
\bea
   \hat{J}^{Z(\gerh^{\bc})} = i \al_+ \partial_z \vphi .
\eea
Combining the compact and non-compact  bosons $\vphi$, $X$
to a $Z(\gerh^{\bc})$-valued complex boson
$\hat {\vphi} =\vphi - iX$,
the TCA of the CG sector
$\{ G^{\pm}_{Z(H^{\bc})}, T_{Z(H^{\bc})}, J_{Z(H^{\bc})} \} $
can be expressed in terms of $\hat{\vphi}$, $\zeta_z$, $\xi$;
\bea
T_{Z(H^{\bc})} &= &
 -\frac{1}{2}: (\partial_z \hat{\vphi}^{\dag} \partial_z \hat{\vphi}) :
 + i \al_-  \rho_{\gerg/\gerh}(\partial_z^2 \hat{\vphi})
              - :(\zeta_z^{Z(\gerh)}, \partial_z \xi^{Z(\gerh)}):,
                     \nonumber \\
G^+_{Z(H^{\bc})} & = & \frac{i}{\sqrt{2}}\,
   ( \xi^{Z(\gerh)},  \partial_z \hat{\vphi})
     + \sqrt{2} \al_-  \rho_{\gerg/\gerh}(\partial_z \xi)  ,\nonumber \\
G^-_{Z(H^{\bc})} &= & \frac{i}{\sqrt{2}}\,
  ( \zeta_z^{Z(\gerh)},  \partial_z \hat{\vphi}^{\dag} )
     +\sqrt{2} \al_-  \rho_{\gerg/\gerh}(\partial_z \zeta_z^{Z(\gerh)}) ,
                                   \nonumber  \\
J_{Z(H^{\bc})} & = & :(\xi^{Z(\gerh)}, \zeta_z^{Z(\gerh)}):
  + i\al_- \rho_{\gerg/\gerh}(\partial_z \hat{\vphi})
  -  i\al_-  \rho_{\gerg/\gerh}(\partial_z \hat{\vphi}^{\dag}) .
\eea
These precisely coincide with those obtained by twisting
the $N=2$ Coulomb gas model\cite{CG}.

{}~

\section{BRST Analysis and the Chiral Primary Ring}
\cleqn
\subsection{The BRST Cohomology on the Semi-Classical Hilbert Space}

{}~

                          Here let us consider the physical states
of the gauge fixed system.
They are characterized by the total BRST-charge
$Q_{\msc{tot}} = Q_{G/H} + Q_{H_0^{\bc}}
+ Q_{Z(H^{\bc})}$ (\ref{totalQ}).
The total Hilbert space $\ca{H}$
is spanned by the state vectors having the form
$| \mbox{WZW} ~ (g)\rangle \otimes |\mbox{WZW} ~ (\rho)\rangle
\otimes | X \rangle
\otimes |\chi \psi \rangle \otimes
|\zeta \xi \rangle$, and the physical Hilbert space $\ca{H}_{\msc{phys}}$
is defined as the $Q_{\msc{tot}}$-BRST cohomology group
in the standard manner;
\be
 \ca{H}_{\msc{phys}} = H^*_{Q_{\msc{tot}}}(\ca{H}).
\label{physical  hilbert space}
\ee
Instead of considering this total cohomology
directly, we shall only  estimate  it ``step by step''. Namely we consider
$  H^p_{Q_{Z(H^{\bc})}} \co H^q_{Q_{H_0^{\bc}}} \co H^r_{Q_{G/H}}(\ca{H}) $,
in order to make the problem simple.
In general this may give  only a subspace of
the precise physical Hibert space
$H^*_{Q_{\msc{tot}}}(\ca{H})$, but, if we can expect
the corresponding spectral sequence  degenerates at  the 2nd order,
it coincides with the physical Hilbert space itself.

First we we shall restrict our attention to the states realized
by the direct products of the primary states of all the dynamical
variables.
In other words we  replace  the total Hilbert space $\ca{H}$
by its ``semi-classical subspace'';
\be
\ca{H}^{\msc{s.c}} \equiv \{ | A\rangle  \in \ca{H} ~;~
  | A \rangle ~\mbox{is~primary} ~\}
\ee
in \eqn{physical hilbert space}.

The most non-trivial part of the cohomology  calculation is
the estimation of $H^r_{Q_{G/H}}(\ca{H}^{\msc{s.c}})$.
It is clearly the  same algebra
as the chiral primary ring in the
$G/H$-Kazama-Suzuki model, which was fully investigated in the papers
\cite{LVW,HT,Gepner}.
In order to present these results it is convenient to introduce
some notations of Lie algebra.
Let $W$ be the Weyl group of $\gerg^{\bc}$.
For any $w \in W$  we set
$\Phi_w \equiv  w(\De^-) \cap \De^+ $
and define $l(w) \equiv \sharp \Phi_w$
(the ``minimal length'' of $w$).
We also set
$W (\gerg / \gerh) \equiv \{ w \in W~;~ \Phi_w \subset
   \De^+_{\germ} \} $.
With these notations $H_{Q_{G/H}}^r (\ca{H}^{\msc{s.c}}) $
is spanned by the following elements;
\be
\ba{l}
 \hat{J}_0^{a_1} \cdots \hat{J}_0^{a_n} | \La , w \rangle_{G/H}
  \otimes \mbox{any state vector of $\rho$, $X$, $\zeta$, $\xi$},  \\
\dsp |\La , w \rangle_{G/H}  \equiv   | \La, w(\La) \rangle_g \otimes
  \prod_{\al \in \Phi_w} \psi^{\al}_0 |0\rangle_{\chi \psi} ,
\ea
\label{La w}
\ee
where $w$ is any element of $W(\gerg /\gerh) $ such that
$l(w) = r$ \cite{LVW,HT,Gepner}.
Notice that $\lb Q_{G/H} , \hat{J} \rb = 0$ holds.
Of course, precisely speaking, \eqn{La w} only expresses
one  representative of the corresponding  cohomology class. One can always
add to it any $Q_{G/H}$-exact term.
\eqn{La w} satisfies the  condition $G^-_{G/H,0} |\Psi \rangle=0$,
besides the BRST invariance $Q_{G/H} | \Psi \rangle =0$.
These states are called as ``chiral primary states'' in SCFT
\cite{LVW,HT,Gepner}, and correspond to ``harmonic cocycles''
in  mathematical terminology.

                               Nextly we should take further
the cohomologies with respect to $Q_{H_0^{\bc}}$ and $Q_{Z(H^{\bc})}$.

                                  Consider the $H_0^{\bc}$-part.
Under the action of $\hat{J}^{\gerh_0}_{0}$(\ref{hatj})
we can extract irreducible
$\gerh_{0}^{\bc}$-modules (with respect to $\hat{J}_0$)
from $H^r_{Q_{G/H}} (\ca{H}^{\msc{s.c}})$;
\be
 H^r_{Q_{G/H}} (\ca{H}^{\msc{s.c}}) =
   \sum_{\La , w} \ca{H}^{G/H}(\La , w) \otimes
   \ca{H}^{\msc{s.c}}_{\rho, X, \zeta \xi} ~ ,
\ee
where $\ca{H}^{\msc{s.c}}_{\rho, X, \zeta \xi}$
denotes the space of the primary
states of $\rho$, $X$, $\zeta \xi$ and
$\dsp \ca{H}^{G/H}(\La , w)  \equiv
  \sum_{\{a_i\}} \bc \hat{J}_0^{a_1} \cdots \hat{J}_0^{a_n}
     | \La , w \rangle_{G/H}$
is the irreducible $\gerh_0^{\bc}$-module
with the highest weight vector $| \La , w \rangle_{G/H}$
having the highest weight
$w * \La|_{\gerh_0^{\bc}}$ \cite{LVW,HT,Gepner}.
Here we introduce the notation;
\be
w * \La = w(\La + \rho_{\gerg}) - \rho_{\gerg} ,
\label{w * La}
\ee
and $|_{\gerh_0^{\bc} }$
denotes the projection to
$\gerh_0^{\bc} \cap \gert^{\bc}$, i.e. the Cartan subalgebra (CSA) of
$\gerh_0^{\bc}$. $\gert^{\bc}$ is the CSA of $\gerg^{\bc}$.
$w * \La|_{\gerh_0^{\bc}}$ is dominant integral
with respect to $\gerh^{\bc}_0$
as is shown from the definition of $W(\gerg/\gerh)$.
Fix one of the pair $(\La, w)$ and consider
$Q_{H_0^{\bc}}$-cohomology on the corresponding space
$\ca{H}^{G/H}_{\La , w} \otimes \ca{H}^{\msc{s.c}}_{\rho, X, \zeta \xi}$,
which is also a $\gerh_0^{\bc}$-module with respect to the total
$\gerh_0$-current
$J^{\gerh_0}_{\msc{tot}} = \hat{J}^{\gerh_0} + J_{\rho} +
J^{\gerh_0}_{\zeta \xi}$
\eqn{jhtot},
and then the desired
$Q_{H_0^{\bc}}$-cohomology is
nothing but the Lie algebra cohomology of $\gerh_0^{\bc}$.
In order to proceed further it is necessary to fix an appropriate
$\gerh_0^{\bc}$-module as the (semi-classical) Hilbert space
of $\rho$. Here we shall take a $\gerh_0^{\bc}$-module with the
highest weight\footnote
  {In the ref.\cite{AGSY} a different choice is taken
  for this representation of $\gerh_0^{\bc}$.
  Because the total BRST cohomology depends on this choice
  the results given in \cite{AGSY} do not fully coincide with
  these we present in this paper.};
\be
\la(\La ,w) \df \overline{w * \La}|_{\gerh_0^{\bc}}
  ~ (\equiv ~ \mbox{the conjugate to $w * \La|_{\gerh_0^{\bc}}$}),
\label{la La w}
\ee
and denote it by $\ca{H}_{\rho}(\la(\La,w))$.
In this choice the $Q_{H_0^{\bc}}$-cohomology  is easily solved;
\be
\ba{l}
H^q_{Q_{H_0^{\bc}}} (\ca{H}^{G/H}(\La , w)
\otimes \ca{H}^{\msc{s.c}}_{\rho, X, \zeta \xi}) \\
{}~~~~~~~~ \cong    H^q (\gerh_0^{\bc};\bc) \otimes
\mbox{Inv}_{\gerh_0^{\bc}} \lb \ca{H}^{G/H}(\La , w)
\otimes \ca{H}_{\rho} (\la(\La, w) ) \rb \otimes
   \ca{H}^{\msc{s.c}}_{X,\zeta^{Z(\gerh^{\bc})}\xi^{Z(\gerh^{\bc})} } .
\ea
\label{eq:coho10}
\ee
This is because $\gerh_{0}^{\bc}$ is semi-simple and
$\ca{H}^{G/H}(\La , w)\otimes \ca{H}^{\msc{s.c}}_{\rho, X, \zeta \xi}$
is finite dimensional.
In the R.H.S of (\ref{eq:coho10}) the first factor
$H^q (\gerh_0^{\bc};\bc)$
corresponds to the contribution of the $\gerh_0^{\bc}$-component
of $\zeta \xi$-ghost system, and the second piece is the
singlet tensors for the global $H_0$-rotations.
$\ca{H}^{\msc{s.c}}_{X,\zeta^{Z(\gerh^{\bc})}\xi^{Z(\gerh^{\bc})} }$
is the semi-classical Hilbert space of $X$ and the
$Z(\gerh^{\bc})$-component of $\zeta \xi$-ghost system.

                           Finally  we should take the cohomology for
the BRST-charge $Q_{Z(H^{\bc})}$.
First let us consider the sector with no $\zeta\xi$-ghost.
Obviously all we have to do is to construct the singlet states for
global $Z(H^{\bc})$-rotation.
We find that any element of the $Q_{G/H}$ and $Q_{H_0^{\bc}}$-cohomology
space above constructed
 has a definte $Z(\gerh^{\bc})$-charge;
 $w * \La|_{Z(\gerh^{\bc})} + \mbox{charge of $X$}$.
Hence the desired result is simple;
if and only if the $Z(H^{\bc})$-charge of $X$ is equal to
the value $- w * \La|_{Z(\gerh^{\bc})}$,
we  get the non-trivial BRST-cohomology.
The sector with the $\zeta \xi$-ghosts is also simple.
We point out the following fact:
Let $\xi^1,\cdots , \xi^l$ be the $Z(\gerh^{\bc})$-components
of the ghost field $\xi$.
Assume $|I \rangle$ satisfies
$J_{\msc{tot},0}^{Z(\gerh)i}|I \rangle = I^i |I \rangle$,
$I^i \neq 0$ for $^{\forall} i \in  S \subset \{1, \ldots , l\}$,
then, for $^{\forall} S' \supset S$,
$\dsp \prod_{i\in S'}\xi_0^i|I \rangle $  is
$Q_{Z(H^{\bc})}$-invariant.
But it  is BRST-trivial except only the case $S = \emptyset $.
In fact we find that
\be
\prod_{i\in S'}\xi_0^i | I \rangle =
Q_{Z(H^{\bc})} \frac{1}{I^{i_0}}\prod_{i\in S'\setminus \{ i_0\}} \xi^i_0
   | I \rangle
\ee
for any $i_0 \in S$.
This observation leads to the fact that
 the $\zeta \xi$-ghost sector is completely factorized
like as the $H^{\bc}_0$-part, namely,
the physical  state with the $Z(H^{\bc})$-ghost number $p$
can be explicitely written as\footnote
   {One might think that, because
   $\xi_0^i = \{ Q_{Z(H^{\bc})} , \al_+ X_0^i \} $ holds,
  any  state including the zero-modes of $\xi$
   becomes BRST-trivial, even if it has the form of  \eqn{xi sector}.
   But, actually it is not the case, since
  the operator $X_0^i$ {\em cannot\/}
  act on the Hilbert space of the states
  possessing  the definite $Z(H^{\bc})$-charge.}
\be
\ba{l}
\dsp \mbox{an element of $H_{Q_{Z(H^{\bc})}}^0 \co H^q_{Q_{H_0^{\bc}}}
   \co H^r_{Q_{G/H}} (\ca{H}^{\msc{s.c}} )$}
    \otimes \prod_{i\in S} \xi_0^i|0\rangle _{\zeta\xi},  \\
S \subset \{ 1, \ldots , l \} ,~~~ \sharp S = p .
\label{xi sector}
\ea
\ee

                      To sum up, the desired physical states
can be written as follows:
Let us denote the cohomology state corresponding to $\La$, $w$ and having
no $\zeta\xi$-ghosts (of both the $H_0^{\bc}$ and the $Z(H^{\bc})$-part)
by $| \La, w \rangle$, which has $l(w)$ as the $\chi\psi$-ghost number
and invariant under the chiral
$H^{\bc} \equiv H_0^{\bc} \times Z(H^{\bc})$-gauge transformations.
The semi-classical physical Hilbert space can be expressed as
\be
\ca{H}^{\msc{s.c}}_{\msc{phys}} = \sum_{\La, w} \bc | \La, w \rangle
  \otimes  \sum_{S}\bc \prod_{i\in S} \xi^i_0 |0\rangle_{\zeta\xi}
    \otimes H^*(\gerh_0^{\bc} ; \bc)   .
\label{sc phys hilbert space}
\ee

                     Now let us turn our interests to the physical
observables. We shall write the ``chiral primary operator''
corresponding to the physical state $|\La, w \rangle$ as $O_{\La,w}(x)$;
\be
  O_{\La ,w}(0)|0\rangle = | \La , w\rangle .
\label{cpo}
\ee
What ring structure do these operators have?
Because it reflects only the local structure of the model,
we may describe it by using some technique of CFT.
Defining its structure constant  by
\be
   O_{\La_1,w_1} O_{\La_2,w_2} = \sum_{\La_3, w_3}
C^{(\La_3,w_3)}_{(\La_1,w_1) \, (\La_2,w_2)}\, O_{\La_3, w_3}
   ~~~(\mbox{modulo BRST-exact terms}),
\ee
we will get the following result;
\be
\ba{l}
  C^{(\La_3,w_3)}_{(\La_1,w_1) \, (\La_2,w_2)}   \propto
   \dsp    F(G_k)^{\La_3}_{\La_1,\La_2} \, \prod_{i=1}^{r}
        F(H_{0,k+g^{\vee}-h_i^{\vee}}^{(i)})^{w_3 *\La_3|_i}
         _{w_1 * \La_1|_i , w_2*\La_2|_i}    \\
 ~~~~~\times \delta(w_1(\La_1)|_Z + w_2 (\La_2)|_Z - w_3 (\La_3)|_Z)\,
          \delta(w_1* 0|_Z + w_2 *0|_Z - w_3 *0|_Z),
\ea
\label{ring}
\ee
where $F(G_k)$, $F(H_{0, k+g^{\vee}-h^{\vee}_i}^{(i)})$ mean
the fusion coefficients of the corresponding current algebras,
and ``delta function'' is defined by
\be
\delta (x) = \left\{
                \ba{ll} 1 & x =0  \\
                        0 & x \neq 0 .
              \ea \right.
\ee
The notations $\La|_i$, $\La|_Z$ mean the orthogonal projections
to $\gerh_0^{(i)\bc}$, $Z(\gerh^{\bc})$ respectively.
The appearance of $F(G_k)^{\La_3}_{\La_1,\La_2} $
in (\ref{ring}) is due to the current algebra $J_g$,
that is, $G_k$-WZW model.
The current algebras $\hat{J}^{\gerh_0} \equiv (\hat{J}^{\gerh_0^{(1)}},
\ldots , \hat{J}^{\gerh_0^{(r)}})$ will give the factor
$\prod_{i=1}^{r}
        F(H_{0,k+g^{\vee}-h_i^{\vee}}^{(i)})^{w_3 *\La_3|_i}
         _{w_1 * \La_1|_i , w_2*\La_2|_i}$
in (\ref{ring}), which will include the contribution
of the $\rho$-sector
($\prod_i^rH_{-(k+g^{\vee}+h_i^{\vee})}^{(i)}$-WZW model).
\footnote{Because the fusion rule of a current algebra
is deeply connected with the structure of null vectors
the negative level current algebra
$H_{0,-(k+g^{\vee}+h_i^{\vee})}^{(i)}$ does not give
stronger condition than its positive level counter part
$H_{0,k+g^{\vee}-h_i^{\vee}}^{(i)}$.}
$\delta(w_1(\La_1)|_Z + w_2 (\La_2)|_Z - w_3 (\La_3)|_Z)$
is due to the conservation of the $Z(\gerh^{\bc})$-charge of $g$-sector.
$\delta(w_1* 0|_Z + w_2 *0|_Z - w_3 *0|_Z)$ reflects the conservation
of the $Z(\gerh^{\bc})$-charge of the $\chi\psi$-sector,
which we can derive by using the identity ;
$\dsp -w* 0 |_{Z} = \{ \rho_{\gerg} - w(\rho_{\gerg})\}|_Z =
\sum_{\al\in \Phi_w} \al |_Z $.
The $Z(\gerh^{\bc})$-charge conservation of the $X$-sector is
automatically ensured by
those of $g$ and $\chi\psi$-sector because of the BRST-invariance.
We also note that the conservation of the N=2 $U(1)$-charge
(the eigenvalue of $J_{G/H,0}$) is included in the above
$Z(\gerh)$-charge consesrvations.

                 The structure constant \eqn{ring}
has a complicated form, but if we only consider
some suitable subring, we can get more simple results.
For example, let us consider the no-ghost sector, i.e.
the physical operators of the form $O_{\La, 1}$, (``$1$'' means
the identitiy in the Weyl group, which trivially belongs
to $W(\gerg/\gerh)$);
\be
C^{(\La_3,1)}_{(\La_1,1) \, (\La_2,1)}   \propto
      F(G_k)^{\La_3}_{\La_1,\La_2}\, \delta(\La_1|_{Z}+ \La_2|_{Z}
                         - \La_3|_{Z}),
\label{ring 2}
\ee
which is the structure constant introduced by Gepner in the case
of $G/H = \bc P^N$ \cite{Gepner 2}.
In particular, in the case of $G/H = G/T$ ($T$ is the maximal torus of $G$),
we get
\be
C^{(\La_3,1)}_{(\La_1,1) \, (\La_2,1)}   \propto
       \delta(\La_1 + \La_2 - \La_3),
\label{ring 3}
\ee
since in this case $Z(H^{\bc}) = T^{\bc}$ holds and $H_0^{\bc}$ is absent.

                In this subsection we only considered
the semi-classical physical
observables which will correspond to the solutions of the
equations of motion ;
$\deebar (\partial g g^{-1})= 0$,
$\deebar \psi = 0$ etc.
But, under some non-trivial background,
i.e. $ \dsp c_1 \equiv \frac{i}{2\pi} \int F(a) \neq 0$,
new observables other than
the above semi-classical ones will appear.
They  may be interpreted as ``instanton-sectors''
which will correspond to the solutions of equations of motion;
$\deebar _a (\partial_a g g^{-1})= 0$,
$\deebar _a \psi = 0$ etc. with $c_1 \neq 0$.
To study these  instanton contributions   we should
take the full Hilbert space into account.

\subsection{The BRST-Cohomology on the Total Hilbert
Space and Spectral Flow}

{}~

                          Let us return to the problem of
solving the BRST-cohomology on the
total Hilbert space.
We will adopt the same strategy as for the semi-classical case.
Namely we will consider
$H^r_{Q_{Z(H^{\bc})}} \co H^q_{Q_{H_0^{\bc}}} \co
H^r_{Q_{G/H}}(\ca{H})$
instead of studying the cohomology
$H^{p+q+r}_{Q_{\msc{tot}}}(\ca{H})$
directly.

{}~

                The results in this section will be
described by using the terminology of
affine Lie algebra.
Besides the notations for affine Lie algebra some mathematical formulae
which we need in this section are summarized in appendix B.
For example we will denote the sets of positive (negative) roots of
the  $\gerg^{\bc}, \gerh_0^{\bc}$-current algebras by
$\hat{\Delta}^+ (\hat{\Delta}^-)$,
$\hat{\Delta}^+_{\gerh_0}
(\hat{\Delta}^-_{\gerh_0})$ respectively.
$\hat{\Delta}^+_{(\gerh_0)}$ consists of elements,
$\alpha~(\alpha \in \Delta^+_{(\gerh_0)})$
and
$\alpha +n \delta~(\alpha \in \Delta_{(\gerh_0)},n \in Z_{>0})$.
$\delta $ is the  generator of imaginary roots.
The modes of the coset,  $H_0^{\bc}$-ghost fields will be
labelled by the elements of
$\hat{\Delta}^+_{\germ}=\hat{\Delta}^+ \setminus \hat{\Delta}^+_{\gerh_0}$,
$\hat{\Delta}_{\gerh_0}
=\hat{\Delta}^+_{\gerh_0} \coprod  \hat{\Delta}^-_{\gerh_0} $.
The (affine) Weyl groups of $\gerg^{\bc}, \gerh^{\bc}_{0}$-current algebras
will be denoted by
$\hat{W}$, $\hat{W}(\gerh_0)$.
Any element $\hat{w}\in \hat{W}$
can be uniquely expressed  as
$t_{\alpha}w ~ (w \in W, \alpha \in Q:$ the root lattice of $\gerg^{\bc}$).
$t_{\alpha}(\alpha \in Q)$ is the ``translation'' by $\alpha$.

                    Firstly we pay attention to the estimation of
$H^r_{Q_{G/H}}$.
This cohomology problem was fully studied in the papers \cite{LVW,HT}.
It was shown that the cohomology elements are labelled by
\bea
(\hat{\Lambda},\hat{w})
\in \hat{P}^k_{+} \times \hat{W}(\gerg /\gerh),
\label{parametrization1}
\eea
where $\hat{P}^k_{+}$ is the set of dominant integral weights
of $\gerg^{\bc}$-current algebra with level $k$.
$\hat{W}(\gerg /\gerh)$
is the subset of $\hat{W}$ which  elements satisfy the condition;
$\Phi_{\hat{w}} \subset \hat{\Delta}^+_{m}$ ,
where we set
$\Phi_{\hat{w}}=\hat{w}(\hat{\Delta}^-) \cap \hat{\Delta}^+$.
$H^r_{Q_{G/H}}(\ca{H})$
can be described as follows;
\bea
H^r_{Q_{G/H}}(\ca{H})
=
\sum_{\hat{\Lambda},\hat{w}}
\ca{H}^{G/H}(\hat{\Lambda},\hat{w})
\otimes
\ca{H}_{\rho, X,\zeta\xi}.
\label{afinQG/H}
\eea
$\hat{w}=t_{\alpha}w
\in \hat{W}(\gerg /\gerh)$
in the R.H.S of (\ref{afinQG/H})
are those elements which satisfy
$r= l(w)-2(\rho, \alpha)$.
$\ca{H}^{G/H}(\hat{\Lambda},\hat{w})$
is spanned by the following vectors
\footnote{
   We label the modes of
   $\chi_{\alpha}(z)=\sum_{n}\chi_{\alpha,n}z^{-n-1}$,
   $\psi^{\alpha}(z)=\sum_{n}\psi^{\alpha}_{n}z^{-n}$
   ($\alpha \in \Delta^+_{\germ}$)
   by $\bm{\psi}_{\hat{\alpha}}$
   ($\hat{\alpha}=\alpha +n \delta \in \hat{\Delta}_{\germ}$),
   \bea
   \bm{\psi}_{\hat{\alpha}=\alpha +n \delta}
   =
   \left\{\begin{array}{ll}
    \psi^{-\alpha}_{n} & \mbox{for}~\alpha \in \Delta^{-}_{\germ} \\
   \chi_{\alpha,n}  & \mbox{for}~\alpha \in \Delta^{+}_{\germ}
   \end{array}\right. \nonumber
    \eea } ;
\bea
&&
\hat{J}^{a_1}_{-m_1} \cdots
\hat{J}^{a_n}_{-m_n}
|\hat{\Lambda},\hat{w} \rangle_{G/H} ~~~~~~
(m_1,\cdots,m_n \in Z_{\geq 0}),\\
&&
|\hat{\Lambda},\hat{w} \rangle_{G/H}=
|\hat{\Lambda},\hat{w}(\Lambda) \rangle_{g} \otimes
\prod_{\hat{\alpha}\in \Phi_{\hat{w}}}
\bm{\psi}_{-\hat{\alpha}}|0 \rangle_{\chi \psi}.
\label{gharmonicc}
\eea
Under the action of the $\gerh_0^{\bc}$-current algebra
$\hat{J}^{\gerh_0}$ (\ref{hatj})
$\ca{H}^{G/H}(\hat{\Lambda},\hat{w})$
is the irreducible $\hat{\gerh}_{0}^{\bc}$-module with the
highest weight vector
$|\hat{\Lambda},\hat{w} \rangle_{G/H}$
which has the weight for the $\gerh_0^{(i)\bc}$-direction;
\be
\hat{w} * \hat{\Lambda}|_{i} + (k+g^{\vee}-h_i^{\vee})\La_0 ~~
(\hat{w}* \hat{\La} \equiv
\hat{w}(\hat{\Lambda}+\hat{\rho}_{\gerg})-\hat{\rho}_{\gerg}) ,
\label{hat w * La}
\ee
where $\hat{\rho}_{\gerg}=\rho_{\gerg}+  g^{\vee} \Lambda_0.$
"$|_i$" means  taking the classical part of $\hat{w}* \hat{\La}$
and then  projecting it to
the $\gerh_0^{(i)\bc}$-component.
$\ca{H}_{\rho,X,\zeta,\xi}$ in the R.H.S of
(\ref{afinQG/H}) is the Hilbert space of
$\rho,X,(\zeta,\xi)$ fields.

                Nextly we should take the cohomology with respect to
$Q_{H^{\bc}_{0}}$ (\ref{qhc}).
For this purpose we should take an appropriate representation
for $\rho$ field.
As is the semi-classical case we may choose,
as the Hilbert space of $\rho$,
the $\hat{\gerh}_{0}^{\bc}$-module which highest weight
is given by
\be
  \hat{\la}^{(i)}(\hat{\La},\hat{w}) \df
     \overline{\hat{w} * \hat{\Lambda}|_{i}} + (-k-g^{\vee}-h^{\vee}_i)\La_0
\label{hat la La w}
\ee
for the $\hat{\gerh}^{(i)\bc}_0$-component.
We write it as
\be
\ca{H}_{\rho}(\hat{\la}(\hat{\La}, \hat{w}) )
=\sum \bc J_{\rho,-m_1}^{a_1}\cdots J_{\rho,-m_n}^{a_n}
\ket{\hat{\la}(\hat{\La}, \hat{w}), \hat{\la}(\hat{\La}, \hat{w})}_{\rho} .
\ee
With this choice
there exist the following  elements of
$H^0_{Q_{H_0^{\bc}}}\co H^0_{Q_{G/H}}(\ca{H})$;
\bea
\mbox{Inv}_{\gerh_0^{\bc}}
\lb \ca{H}^{G/H~(\msc{s.c})}(\hat{\Lambda},\hat{w})
\otimes
\ca{H}^{(\msc{s.c})}_{\rho}
  (\hat{\la}(\hat{\La}, \hat{w})) \rb  \otimes |0 \rangle_{\zeta \xi},
\label{afinQH0QGHco}
\eea
where
\be
\ba{lll}
\ca{H}^{G/H~(\msc{s.c})}(\hat{\Lambda},\hat{w}) &=&
\sum \bc \hat{J}^{a_1}_{0}\cdots \hat{J}^{a_n}_{0}
|\hat{\Lambda},\hat{w}\rangle_{G/H},  \\
\ca{H}^{(\msc{s.c})}_{\rho}
  (\hat{\la}(\hat{\La}, \hat{w}))
&=&\sum
\bc J^{a_1}_{\rho , 0}\cdots J^{a_n}_{\rho , 0}
\ket{\hat{\la}(\hat{\La}, \hat{w}),\hat{\la}(\hat{\La}, \hat{w})}_{\rho}
\ea
\ee
are the semi-classical Hilbert space of
$\ca{H}^{G/H}(\hat{\Lambda},\hat{w})$,
$\ca{H}^{(\msc{s.c})}_{\rho}
  (\hat{\la}(\hat{\La}, \hat{w}))$
respectively.
The global $\gerh^{\bc}_{0}$-invariance in
(\ref{afinQH0QGHco}) ensures the $Q_{H^{\bc}_{0}}$-invariance
of the state
since the ($\zeta,\xi$)-vacuum state $|0 \rangle_{\zeta \xi}$ is
characterized by the conditions\footnote
   {We label the modes of
   $\zeta(z)=\sum_{n}\zeta_nz^{-n-1},\xi(z)=\sum_{n}\xi_{n}z^{-n}$ by
   $\bm{\zeta}_{\alpha+n \delta}$, $\bm{\xi}_{\alpha +n \delta}$
   $(\alpha \in \Delta_{\gerh_0})$,
   $\zeta_n^{\gert}$ and $\xi^{\gert}_n$;
   \bea
   \bm{\zeta}_{\alpha +n \delta}
   =
   \left\{\begin{array}{ll}
   \zeta_{\alpha ,n} & \mbox{for}~\alpha \in \Delta^{+}_{\gerh_0} \\
   \zeta^{-\alpha}_{n}  & \mbox{for}~\alpha \in \Delta^{-}_{\gerh_0}
   \end{array}\right.~~,~
   \bm{\xi}_{\alpha +n \delta}
   =
   \left\{\begin{array}{ll}
   \xi_{\alpha,n} & \mbox{for}~\alpha \in \Delta^{+}_{\gerh_0} \\
   \xi^{-\alpha}_{n}  & \mbox{for}~\alpha \in \Delta^{-}_{\gerh_0}
   \end{array}\right. \nonumber
   \eea
   }:
$\bm{\zeta}_{\alpha +n \delta}|0\rangle_{\zeta \xi}
=(h,\zeta^{\gert}_{n })|0\rangle_{\zeta \xi}=0$
( for $n \geq 0, \alpha \in \Delta_{\gerh_0}, h \in \gert)$,
$\bm{\xi}_{\alpha +n \delta}|0\rangle_{\zeta \xi}
=(h,\xi^{\gert}_n)|0\rangle_{\zeta \xi}=0$
(for $n > 0, \alpha \in \Delta_{\gerh_0}, h \in \gert)$.
$\gert$ is the CSA of $\gerg$.
Lastly we should take the cohomology with respect to
$Q_{Z(H^{\bc})}$ (\ref{qhc}).
It will be achieved by imposing the global
$Z(\gerh^{\bc})$-invariance on the state (\ref{afinQH0QGHco}),
which gives rise to the following element that belongs to
$H^0_{Q_{Z(H^{\bc})}} \co H^0_{Q_{H_0^{\bc}}} \co
H^r_{Q_{G/H}}(\ca{H})$;
\be
\ket{\hat{\La}, \hat{w}} \df
\mbox{Inv}_{\gerh_0^{\bc}}
\lb
\ca{H}^{G/H~(\msc{s.c})}(\hat{\Lambda},\hat{w})
\otimes
\ca{H}^{(\msc{s.c})}_{\rho}(\hat{\la}(\hat{\La}, \hat{w})) \rb
\otimes
\ket{-\hat{w}*\hat{\Lambda}|_Z}_{X}
\otimes
|0 \rangle_{\zeta \xi}.
\label{afinQQQco}
\ee
Here ``$|_Z$'' stands for the similar meaning as $|_i$, i.e.
the $\hat{w}*\hat{\La}|_Z$ is the $Z(\gerh^{\bc})$-projection
of the classical part of $\hat{w}*\hat{\La}$.

{}~


                         To proceed further let us
introduce  a  powerful tool - the ``spectral
flow''  $\ca{U}$ \cite{LVW,HT}, which is a family
of infinite  symmetry transformations
of our topological model in the sense that they make the BRST charge
$Q_{\msc{tot}}$ (\ref{totalQ}) invariant;
\bea
\ca{U}\, Q_{\msc{tot}}\, \ca{U}^{-1}=Q_{\msc{tot}},
\label{UQ=QU}
\eea
and that they change the background gauge fields
of the gauge fixed model appropriately.
The second point will be discussed in the next section.
These transformations are essentially induced from the ``translations''
in the (affine) Weyl group of  $\gerg^{\bc}$-current algebra.

                        Let us describe them explicitly.
For this purpose we introduce the following subset of
the weight lattice $P$ of $\gerg^{\bc}$
which labels the spectral flow $\ca{U}$;
\bea
\ca{P}(\gerg/\gerh)=
\{ ~
\gamma \in P ~;~
^{\exists}\sigma \in W(\gerh_0) ~\mbox{s.t} ~
\sigma(C_{0,\gerh_0}^{\msc{aff}}+\gamma)=C_{0,\gerh_0}^{\msc{aff}}
{}~ \},
\label{P(g/h)}
\eea
where $W(\gerh_0)$ is the Weyl group of $\gerh_0$
and $C_{0, \gerh_{0}}^{\msc{aff}}$ is the subdomain of
$\gert^{*}$
which contains the Weyl alcove of $\gerh_0$;
\bea
C_{0,\gerh_0}^{\msc{aff}}=
\{ ~
u \in \gert^* ~;~
(u,\, ^{\forall}\alpha_{l_i}^{(i)})\geq 0,(u,\, ^{\forall}\theta^{(i)})\leq 1
 ~ \}.
\label{alcove}
\eea
$\alpha_{l_i}^{(i)}$
$(1 \leq l_i \leq \mbox{rank}\gerh^{(i)\bc}_{0})$
are the simple roots of $\gerh_{0}^{(i)\bc}$
$(1\leq i \leq r)$,
and $\theta^{(i)}$ is the maximal root of $\gerh_{0}^{(i)\bc}$
$(1\leq i \leq r)$.
With these definitions it is clear that
$\sigma \in W(\gerh_0)$
in (\ref{P(g/h)}) is uniquely
determined for $\gamma \in P$
if it exists, and we shall denote it by $\sigma_{\gamma}$.
We further introduce the notation;
\be
 \hat{w}(\gamma) = \sigma_{\gamma} t_{\gamma}
\label{w gamma}
\ee
for any element of $\gamma \in \ca{P}(\gerg/\gerh)$.
The spectral flow $\ca{U}_{\gamma}$
will be defined as the action of $\hat{w}(\gamma)$.

                              We are now in a position to write
the defintion of spectral flow.
Under the action of $\ca{U}_{\gamma}$ ($\gamma \in P(\gerg/\gerh)$)
the $\gerg^{\bc}$-current algebra $J_{g}$ and the ghost fields
($\chi, \psi$) should be transformed into
$\ca{U}_{\gamma}\, J_g \, \ca{U}_{\gamma}^{~-1},
\ca{U}_{\gamma}\,  \chi \, \ca{U}_{\gamma}^{~-1},
\ca{U}_{\gamma}\,  \psi \, \ca{U}_{\gamma}^{~-1}$;
\bea
\ca{U}_{\gamma}\, J_{g,\alpha}(z) \, \ca{U}_{\gamma}^{~-1}
&=&J_{g,\sigma_{\gamma}(\alpha)}(z)z^{-(\alpha,\gamma)} \nonumber \\
\ca{U}_{\gamma}\,J_{g}^{\alpha}(z) \, \ca{U}_{\gamma}^{~-1}
&=&J_{g}^{\sigma_{\gamma}(\alpha)}(z)z^{(\alpha,\gamma)}  \label{sflow g}\\
\ca{U}_{\gamma}\, (h,J_g)(z) \, \ca{U}_{\gamma}^{~-1}
&=&(\sigma_{\gamma}(h),J_g)(z)
   -k\langle \gamma, h\rangle \delta_{n,0} \nonumber \\
\ca{U}_{\gamma}\, \chi_{\alpha}(z) \, \ca{U}_{\gamma}^{~-1}
&=&\chi_{\sigma_{\gamma}(\alpha)}(z)z^{-(\alpha,\gamma)} \nonumber \\
\ca{U}_{\gamma}\,\psi^{\alpha}(z) \, \ca{U}_{\gamma}^{~-1}
&=&\psi^{\sigma_{\gamma}(\alpha)}(z)z^{(\alpha,\gamma)}  \label{sflow chi psi}
\eea
Moreover, by expressing $\hat{w}(\gamma)$ as
$\hat{w}(\gamma)=\hat{\tau}_{\gamma}\hat{\om}_{\gamma}$
$\in D \sdprod \hat{W}$
($\hat{\om}_{\gamma}\in \hat{W}$,
$\hat{\tau}_{\gamma}\in D$ (=
the group of  extended Dynkin diagram automorphisms
of $\gerg^{\bc}$)),
the transformations of the primary states are given by
\bea
\ca{U}_{\gamma}|\hat{\Lambda},\hat{\Lambda}\rangle_{g}
&=&
|\hat{\tau}_{\gamma}(\hat{\Lambda}),
\hat{w}(\gamma)(\hat{\Lambda})\rangle_{g} \\ \nonumber
\ca{U}_{\gamma}|0\rangle_{\chi \psi}
&=&
\prod_{\hat{\alpha}\in \Phi_{\hat{w}(\gamma)}}
\bm{\psi}_{-\hat{\alpha}}|0\rangle_{\chi \psi}.
\label{specgghost}
\eea

               From the transformation rule (\ref{specgghost})
the state
$|\hat{\Lambda},\hat{w}\rangle_{G/H}$ (\ref{gharmonicc})
will be transformed into another physical state
$|\hat{\Lambda'},\hat{w}'\rangle_{G/H}$ by $\ca{U}_{\gamma}$;
\bea
\ca{U}_{\gamma}|\hat{\Lambda},\hat{w}\rangle_{G/H}&=&
|\hat{\Lambda'},\hat{w}'\rangle_{G/H} \nonumber \\
&&
\left\{  \begin{array}{l}
      \hat{\Lambda'}=\hat{\tau}_{\gamma}(\Lambda)\in \hat{P}^k_{+}
  \\
      \hat{w'}=\hat{w}(\gamma)\hat{w}\hat{\tau}_{\gamma}^{-1}
          \in \hat{W}(\gerg/\gerh).
         \end{array}
\right.
\label{specG/H}
\eea
Remark that this $\hat{w}'$ is indeed an element of $\hat{W}(\gerg/\gerh)$.
(See the proposition \ref{prop 3} in appendix B.)
In the standpoint of the  coset CFT it is claimed that any states
$|\hat{\Lambda},\hat{w}\rangle_{G/H}$ which are transformed into
each other by some spectral flow should be identified
\cite{LVW,HT}.
Because there exists the isomorphism between
$\ca{P}(\gerg/\gerh)$ and $P/Q(\gerh_0)$;
\bea
\ca{P}(\gerg/\gerh)\ni \gamma
\longmapsto \mbox{[} \gamma \mbox{]} \in P/Q(\gerh_0),
\label{iso}
\eea
the equivalent class of $H_{Q_{G/H}}(\ca{H})$
(\ref{afinQG/H}) under the identification (\ref{specG/H})
can be labelled by
$\dsp \frac{\hat{P}^k_{+}\times \hat{W}(\gerg/\gerh)}{P/Q(\gerh_0)} \equiv
\frac{\hat{P}^k_{+}\times W(\gerg/\gerh)}{D}$ \cite{LVW,HT}.
Note that $\hat{W}(\gerg/\gerh)\cong W(\gerg/\gerh) \times Q/Q(\gerh_0)$
(see the propositon \ref{prop 4} and the propositon \ref{prop 1}
in appendix B) and $D \cong P/Q$.

                  Nextly we will describe the action of $\ca{U}_{\gamma}$
($\gamma \in \ca{P}(\gerg/\gerh)$) on the $H^{\bc}/H$-WZW sector
(the $\rho$-sector) and the $H^{\bc}$-ghost sector.
The $\gerh_0^{\bc},Z(\gerh^{\bc})$-current algebras
$J_{\rho}$, $J_{X}$ and the ghost fields ($\zeta$, $\xi$)
are transformed into
$\ca{U}_{\gamma}\, J_{\rho}\, \ca{U}_{\gamma}^{~-1},
\ca{U}_{\gamma}\, J_{X}\, \ca{U}_{\gamma}^{~-1}$
and ($\ca{U}_{\gamma}\, \zeta\, \ca{U}_{\gamma}^{~-1},
\ca{U}_{\gamma}\, \xi\, \ca{U}_{\gamma}^{~-1}$).
Their explicit forms are given
in  \eqn{sflow jrho}, \eqn{sflow jX}, \eqn{sflow zeta xi 1},
\eqn{sflow zeta xi 2} in appendix B.
{}From the definition \eqn{w gamma}
$\hat{w}(\gamma)$ acts on $\gerh_{0}^{\bc}$
as an extended Dynkin diagram automorphism of
$\gerh_0^{\bc}$.
We denote its action as $\hat{\tau}^{\gerh_0}_{\gamma}$
$\in D(\gerh_0)$(= the group of the extended Dynkin diagram automorphisms
of $\gerh_0^{\bc}$).
Then $\ca{U}_{\gamma}$ transforms the primary  states into
\bea
\ca{U}_{\gamma}|\hat{\lambda},\hat{\lambda}\rangle_{\rho}&=&
|\hat{\tau}^{\gerh_0}_{\gamma}(\hat{\lambda}),
\hat{\tau}^{\gerh_0}_{\gamma}(\hat{\lambda})
\rangle_{\rho} \\
\label{specH0}
\ca{U}_{\gamma}|\beta \rangle_{X}&=&
|\beta -(k+g^{\vee})\gamma \rangle_{X} \\
\label{specZH}
\ca{U}_{\gamma}|0 \rangle_{\zeta\xi}&=&
|\hat{\tau}^{\gerh_0}_{\gamma}\rangle_{\zeta \xi},
\label{specHghost}
\eea
where the state $|\hat{\tau}^{\gerh_0}_{\gamma}\rangle_{\zeta \xi}$
is defined by
$\dsp \prod_{\alpha \in \Delta_{\gerh_0}^+}
\bm{\zeta}_{-\hat{\tau}_{\gamma}^{\gerh_0}(\alpha)}
\bm{\xi}_{\alpha}|0\rangle_{\zeta \xi}$.
Especially the ($\zeta,\xi$) Fock vacuum can be written
as $|\hat{\tau}^{\gerh_0}_{\gamma}=\mbox{id} \rangle_{\zeta \xi}$.

Since the BRST cohomology state (\ref{afinQQQco}) is composed of
the primary states of all the sectors or their descendents,
its transformation rule under the actions of
spectral flows can be derived from the above formulae.
Moreover, because of the property (\ref{UQ=QU}),
this transformed state is  also
BRST-invariant (and not BRST-trivial).
If taking care of the transformation
rules of the $\rho$-sector (\ref{specH0})
and the $H^{\bc}$-ghost sector (\ref{specHghost}),
we find that the transformed state does not necessarily
have the form of \eqn{afinQQQco}.
Namely we will  obtain new cohomology classes
by making the spectral flows
act on \eqn{afinQQQco}.
We rewrite  the BRST-cohomology state \eqn{afinQQQco} as
$|\hat{\Lambda},\hat{w},\hat{\tau}^{\gerh_0}=\mbox{id} \rangle$.
Then $\ca{U}_{\gamma}$ transforms this states into
$\ca{U}_{\gamma}|\hat{\Lambda},\hat{w},\mbox{id} \rangle$, which
we denote by
$|\hat{\Lambda'},\hat{w'},\hat{\tau}^{\gerh_0}_{\gamma} \rangle$.
$\hat{\Lambda'},\hat{w'}$ are those given in (\ref{specG/H}).
In this way we have obtained  a family of
the physical states labelled by
\bea
(\hat{\Lambda},\hat{w},\hat{\tau}^{\gerh_0})
\in \hat{P}^k_{+}\times \hat{W}(\gerg/\gerh)\times D(\gerh_0).
\label{labelobservable}
\eea
The actions of the spectral flows are closed among them.
This is because the flows have the following property:
For $\gamma_1,\gamma_2 \in \ca{P}(\gerg/\gerh),$
\bea
\ca{U}_{\gamma_1} \,  \ca{U}_{\gamma_2}
&=& \ca{U}_{\gamma_2}\,  \ca{U}_{\gamma_1} \\ \nonumber
&=& \ca{U}_{\gamma_1 \scdplus \gamma_2},
\label{commring}
\eea
where $\gamma_1 \dplus \gamma_2 \in \ca{P}(\gerg/\gerh)$
is defined via the isomorphism (\ref{iso});
\bea
P/Q(\gerh_0)\ni \mbox{[}\gamma_1+\gamma_2 \mbox{]}
\longmapsto  \gamma_1 \dplus \gamma_2 \in \ca{P}(\gerg/\gerh).
\eea
For $\gamma \in \ca{P}(\gerg/\gerh)$,
we also define $\dminus \gamma \in \ca{P}(\gerg/\gerh)$
by the image of $\mbox{[} \gamma \mbox{]}$.


{}~

                              We  denote the physical observable
corresponding to the physical state
$\ket{\hat{\La}, \hat{w}, \hat{\tau}^{\gerh_0}}$
by $O_{\hat{\La}, \hat{w}, \hat{\tau}^{\gerh_0}}$
as in the semi-classical case \eqn{cpo}.
The ring structure of these operators will be
given by the generalization of that of the semi-classical
operators \eqn{ring};
\be
\ba{llll}
  C^{(\hat{\La}_3,\hat{w}_3,\hat{\tau}_3^{\gerh_0})}
  _{(\hat{\La}_1,\hat{w}_1,\hat{\tau}_1^{\gerh_0}) \,
    (\hat{\La}_2,\hat{w}_2,\hat{\tau}_2^{\gerh_0})}
   &\propto&  \dsp    F(G_k)^{\hat{\La}_3}_{\hat{\La}_1,\hat{\La}_2} \,
     \prod_{i=1}^{r}
        F(H_{0,k+g^{\vee}-h_i^{\vee}}^{(i)})
      ^{\hat{\la}_3^{(i)}}_{\hat{\la}_1^{(i)}, \hat{\la}_2^{(i)}}    \\
&&     ~~~ \times \delta(\hat{w}_1(\hat{\La}_1)|_Z
    + \hat{w}_2 (\hat{\La}_2)|_Z - \hat{w}_3 (\hat{\La}_3)|_Z)  \\
&&     ~~~ \times \delta(\hat{w}_1*0|_Z + \hat{w}_2 *0|_Z
          - \hat{w}_3 *0|_Z) \\
&&     ~~~ \times  \delta(\hat{\tau}^{\gerh_0}_1\,\hat{\tau}^{\gerh_0}_2 \,
   \hat{\tau}^{\gerh_0-1}_3 ),
\ea
\label{ring 4}
\ee
where $\hat{\la}^{(i)}_j \equiv \hat{\la}^{(i)}(\hat{\La}_j, \hat{w}_j)$
and $\delta(\hat{\tau}^{\gerh_0}_1\,\hat{\tau}^{\gerh_0}_2 \,
   \hat{\tau}^{\gerh_0-1}_3 )$ stands for
the ''delta function on $D(\gerh_0)$'' defined by
\be
\delta(\hat{\tau}^{\gerh_0})=
\left\{
   \ba{ll}
        1 & \hat{\tau}^{\gerh_0}= \mbox{id} \\
        0 & \hat{\tau}^{\gerh_0}\neq \mbox{id} .
  \ea
\right.
\ee

                   It is an important question
whether the field  identification
by the spectral flow is compatible with this ring structure \eqn{ring 4}.
We rewrite the   physical operators as  $ O_a $ $(a \in I)$.
$I$ is the index set  (\ref{labelobservable}).
Define the action of spectral flow on
$\{ O_a \}_{a \in I}$ by the standard field-state correspondence,
that is, the operator $\ca{U}_{\gamma} (O_a) \equiv
O_{\gamma \cdot a}$ is given by
\be
\ca{U}_{\gamma} (O_a) |0\rangle =
     \ca{U}_{\gamma } |a \rangle ,
\label{defofobservable}
\ee
where $| a \rangle $ is the physical state corresponding to $O_a$.
Then we can show the following identity with respect
to the structure constant $C^c_{ab}$;
\be
  C^{(\gamma_1 \scdplus  \gamma_2) \cdot c}
    _{\gamma_1 \cdot a,  \gamma_2  \cdot b}
     = C^c_{ab},
\label{ring sflow}
\ee
or equivalently,
\be
   \ca{U}_{\gamma_1}(O_a) \ca{U}_{\gamma_2} (O_b)
    = \sum_{c \in I} C^c_{ab}\,
   \ca{U}_{ \gamma_1 \scdplus \gamma_2 }(O_c).
\label{ring sflow 2}
\ee
This can be proved by observing
the explicit form of $C^c_{ab}$ \eqn{ring 4}.
The most non-trivial part of the proof  is the following relation for
the $G$-WZW sector;
\be
   F(G_k)^{\hat{\tau}_{\gamma \scdplus \gamma'}
  (\hat{\La}_3)}_{\hat{\tau}_{\gamma}(\hat{\La}_1),
            \hat{\tau}_{\gamma'}(\hat{\La}_2)}
= F(G_k)^{\hat{\La}_3}_{\hat{\La}_1,\hat{\La}_2} ,
\label{fusion}
\ee
and the similar relations of the $H_0^{(i)}$-parts.
This relation \eqn{fusion} is  shown by making
use of the Verlinde's formula on the fusion coefficients
\cite{Verlinde} and the modular transformation properties
of the affine characters. (Refer for example
\cite{HT,Bernard,KW}.)
In the next section we will derive this formula \eqn{ring sflow 2}
from  a more physical viewpoint, that is, as a direct consequence
of the topological invariance of the system.

                           The identity \eqn{ring sflow} implies that
the field identifications are consistent with
the ring structure of the BRST-cohomology.
Especially if we set
\be
\lb O_a \rb = \{ ~ \ca{U}_{\gamma} (O_a) ~; ~^{\forall} \gamma
                \in \ca{P}(\gerg/\gerh) ~\},
\label{identify}
\ee
we can consistently introduce a product  on them by
\be
\lb O_a \rb \lb O_b \rb \df  \lb  O_a O_b \rb
\ee
without depending on the choice of representatives.
Hence, on the level of local properties of our model,
these field identifications  completely work  and
one may always extract at most finite physical
degrees of freedom.

                     However, once we turn our attention
to the global structure of the
system, we will face with somer subtle problem.
Especially, if we calculate the correlation functions
with some fixed back-ground topology
(Euler number and Chern classes in our case),
we will find that
\be
\left \langle \ca{U}_{\gamma} (O_a)\, O_b\, \ldots\,
  O_c \right\rangle_{g, c_1} \neq
\left \langle O_a\, O_b \, \ldots \, O_c \right\rangle_{g, c_1} ,
\ee
because $O_a$ and $\ca{U}_{\gamma}(O_a)$ have different
ghost numbers although they have the same $N=2$ $U(1)$-charges.
In the next section we  will discuss this problem in detail.

{}~

\section{Correlation Functions of Physical Observables and
a Geometrical Interpretation of Spectral Flow}
\cleqn

{}~

                                      In this section we will study
the correlators of the physical observables
constructed in section 3,
( the correlators among the 0-form components).
We shall perform this estimation with the topology fixed,
that is, the genus $g$ of $\Sigma$
and the ``vector of Chern numbers"
$\dsp c_1 \equiv \frac{i}{2\pi} \int F(a)$ fixed to be some definite values.
For the genus $g$, the higher genus correlation
functions ($g\geq 2$) always become  zero.
This is a simple result obtained from  counting of the  anomaly
for the  $N=2$ $U(1)$-charge
(\ref{jgh})(of the Kazama-Suzuki sector).
Because  all the physical observables
obtained above have non-negative $N=2$ $U(1)$-charges,
while its back-ground charge is
equal to $Q_{\msc{KS}}(1-g)$, which is negative when $g \geq 2$.
($\dsp Q_{\msc{KS}}=
\dimn \germ_+ - \frac{4}{k+g^{\vee}} \rho_{\gerg/\gerh}^2 > 0$.)
We shall consider the genus 0 case. We write
the correlator on sphere as $\langle ~ \cdots~ \rangle_{c_1}$.
We should manifestate the domain in which $c_1$ takes its value.
Notice that $c_1 $ must belong
to the weight lattice $P$.\footnote
   {Precisely speaking, $c_1 \in \nu^{-1}(P)$,
  where $\nu ~:~ \gert^{\bc}~\rarr ~ \gert^{\bc*}$ is the isomorphism
  defined by the inner product. (See appendix B.) But we shall here omit it
 to avoid the complexity of notations.}
The $Z(\gerh^{\bc})$-components of $c_1$ will play as topological
invariants, while the $\gerh_0$-components can be changed
by chiral gauge transformations.
These  gauge degrees of freedom correspond to the shifts
by the  root lattice
$Q(\gerh_0)$ of $\gerh_0$.
Hence  we can assume  $c_1 $ takes its value in
$\ca{P}(\gerg/\gerh)$
(\ref{P(g/h)}),
which is isomorphic to the quotient lattice
$P/Q(\gerh_0)$.

{}~

                        Let us study the correlation function
in the operator formalism.
In this approach the correlation function may be regarded as the
matrix element under the following configuration of the back-ground
gauge field $a=(a_0, i \omega)$
(\ref{bggauge});
\bea
F(a) = - \pi c_1 \de^{(2)}(z-\infty) d \bar{z} \wedge dz.
\label{background1}
\eea
We normalize the delta function as
$\dsp \int \de^{(2)}(z) d \bar{z} \wedge dz=2 i$.
So we can define the correlator by
\be
\left \langle O_{a_1}(x_1) \, O_{a_2}(x_2)\, \ldots
O_{a_3}(x_3) \right \rangle_{c_1} =
 \langle 0 , c_1 | \ca{O}_{\xi}(x_0) \, O_{a_1}(x_1) \, O_{a_2} (x_2) \,
\ldots O_{a_n}(x_n) | 0\rangle .
\label{cor op}
\ee
The in-vacuum
$|0\rangle$
has no back-ground charge. The out-vacuum
$\langle 0, c_1|$
is the BRST-invariant state having the suitable back-ground charges
corresponding to $c_1$.
This is an analoguous situation as that in the Coulomb gas realization
of CFT \cite{DF}.
$\ca{O}_{\xi}(x)$ denotes the BRST-invariant operator which state is
$\dsp \prod_{a \in \gerh}\xi^a_0 | 0\rangle_{\zeta\xi} \otimes
|0\rangle_{\msc{others}}$.
This operator should be inserted
in order to cancel the $\zeta\xi$-ghost number
anomalies.
In the following we may omit to write it explicitly.
We also notice that, since our EM tensor is BRST-exact
(\ref{exacttotT}),
the correlation function (\ref{cor op}) does not depend on the
operator insertion points. So we may also omit to write these
insertion points.

                     The back-ground charges in (\ref{cor op})
can be determined from the
estimations of the ghost number or the chiral anomalies.
Let $\dsp N_{\chi\psi}^{(\al)} \equiv \frac{1}{2\pi i} \oint dz
  : \psi^{\al}\chi_{\al}:~(\al \in \De^+_{\germ})$,
$N_{\zeta\xi}^{(\al)}~(\al \in \De_{\gerh_0})$, and
$N^{(j)}_{\zeta\xi}~(j = 1,\ldots , \mbox{rank}\gerg)$
be the ghost number operators for
$(\chi,\psi),(\zeta,\xi)$-ghost sectors.
The out-vacuum
$\langle 0, c_1|$ should be characterized by the conditions;
\bea
\langle 0, c_1| N^{(\al)}_{\chi \psi}
& = & \langle 0, c_1| \{ 1 + (\al , c_1) \} ,
\label{bg charge chi psi}\\
\langle 0, c_1| N^{(\al)}_{\zeta \xi}
& = & \langle 0, c_1| \{ 1 + (\al , c_1) \} ,
\label{bg charge zeta xi}\\
\langle 0, c_1| N^{(j)}_{\zeta \xi}
& = & \langle 0, c_1|  ,
\label{bg charge zeta xi 2}\\
\langle 0 , c_1| (J_{g,0} , h )
& = & \langle 0, c_1| \{ -k \langle c_1 , h\rangle \} ,~~~
              (\forall h \in \gert^{\bc})
\label{bg charge g}\\
\langle 0 , c_1| (J^{(i)}_{\rho,0} , h )
& = & \langle 0, c_1| \{ (k+g^{\vee}+h^{\vee}_i) \langle c_1 , h\rangle \} ,~~~
              (\forall h \in \gert(\gerh_0^{(i)})^{\bc})
\label{bg charge rho}\\
\langle 0, c_1| ( J_{X,0} , h)
& = & \langle 0, c_1| \{ 2 \langle \rho_{\gerg/\gerh}, h\rangle
    + (k+g^{\vee}) \langle c_1 , h\rangle \},~~~
         (\forall h \in Z(\gerh^{\bc})).  \nonumber \\
&&  ~~~~~
\label{bg charge X}
\eea
It is worthwhile to notice that the relation,
\bea
\langle 0, c_1| J_{G/H,0}  =
      \langle 0, c_1| Q_{\msc{KS}},
\eea
holds {\em without depending on $c_1$\/}, which
suggests some geometrical meaning of the Kazama-Suzuki
$U(1)$-current $J_{G/H}$
(\ref{jgh}).
The charge of it generates a global
$U(1)$-rotation  on both the bosonic part $g$ (and necessarily $X$)
and the fermionic part $\chi\psi$ along
the direction which does not depend on $c_1$.

                        In order to give the explicit form
of the out-vacuum state
we first notice that the state
$\langle 0, c_1=0|$
still has the non-zero back-ground charges which are due to
the Fegin-Fucks term
$\dsp \sim \int \rho_{\gerg/\gerh}(X) R(\bm{g})$ in $S_X$
(\ref{action X}).
This out-vacuum $\langle 0,c_1=0 |$ is realized by
\bea
\ba{lll}
\langle 0, c_1 = 0 |& =& \dsp
\langle k\La_0 , k\La_0|_g
    \otimes \bigotimes_{i=1}^r
    \langle -(k+g^{\vee}+h^{\vee}_i) \La_0, -(k+g^{\vee}+h^{\vee}_i) \La_0
    |_{\rho^{(i)}}  \\
 & & \dsp ~~~~~~ \otimes \langle 2\rho_{\gerg/\gerh}|_X
           \otimes \langle 0|_{\chi\psi} \prod_{\al \in \De_{\germ}^+}
                                                 \chi_{\al,0}
           \otimes \langle 0|_{\zeta\xi} \prod_{a \in \gerh}\zeta_0^a  \\
&\equiv & \langle k\La_0 , w_0 ,0 |   ,
\ea
\label{out 1}
\eea
in  the notation of the previous section:
$(k\La_0 , w_0, \mbox{id}) \in \hat{P}_+^k
\times \hat{W}(\gerg/\gerh) \times D(\gerh_0)$.
$w_0$ is the element having the maximal length in $W(\gerg/\gerh)$.
For the description of
$\langle 0,c_1|$ with general values of $c_1$
we utilize the relation;
\be
\langle 0, c_1| \ca{U}_{\gamma}
  = \langle 0, c_1 \dplus  \gamma | ,
\label{sflow out}
\ee
which follows from the facts that
the spectral flow $\ca{U}_{\gamma}$ preserves the BRST-invariance
and that the state $\langle 0,c_1|\ca{U}_{\gamma}$ satisfies the same
conditions (\ref{bg charge chi psi})-(\ref{bg charge X}) as those
of $\langle 0, c_1\dplus \gamma | $.
This identity (\ref{sflow out}) is siginificant from the view of
topological theory because it enables us to interpret the spectral flows
as the transformations which connect the sectors with different
Chern numbers, that is, the different instanton sectors.
By applying the equality (\ref{sflow out}) to the state
$\langle 0,c_1=0|$ (\ref{out 1}) we can obtain the following
general expression of $\langle 0,c_1|$ :
\be
\ba{lll}
\langle 0, c_1 = \gamma |& = &\langle 0,c_1=0 | \ca{U}_{\gamma} \\
    &= &  \langle \hat{\tau}^{-1}_{\gamma}(k\La_0),
   \hat{w}(\gamma)^{-1}w_0\hat{\tau}_{\gamma},
    \hat{\tau}^{\gerh_0 -1}_{\gamma}| ,
\ea
\label{out 2}
\ee
where $ \hat{\tau}_{\gamma}$, $ \hat{w}(\gamma) $ and
$ \hat{\tau}^{\gerh_0}_{\gamma} $
are, as in the previous section,
the elements uniquely defined by $ \gamma $.

                    To begin with, we shall consider the one point function.
There exists a unique physical operator
$O^{c_1}_{\msc{max}}(x)$ defined by the condition
\be
  \left \langle O^{c_1}_{\msc{max}} \right\rangle_{c_1} =1 .
\label{max}
\ee
This is the operator which state is dual to
the out-vacuum $\langle 0,c_1=\gamma|$ (\ref{out 2}),
that is, the state
$|\hat{\tau}^{-1}_{\gamma}(k\La_0),
   \hat{w}(\gamma)^{-1}w_0\hat{\tau}_{\gamma},
        \hat{\tau}^{\gerh_0-1}_{\gamma}  \rangle$.
It has the maximum U(1)-charge of our model, which is equal to
$Q_{\msc{KS}}$.
For example
$O^{c_1=0}_{\msc{max}}$
is the operator corresponding to
$ | k\La_0 , w_0 , \mbox{id} \rangle $.
It includes the term
$\dsp \prod_{\al \in \De_{\germ}^+} \psi_0^{\al}|0 \rangle _{\chi\psi}$
for the $\chi\psi$-sector.
This makes us interpret $O_{\msc{max}}^{c_1=0}$
as the top cohomology class  of $H_{\msc{DR}}^*(G/H)$ defined by
the volume element of $G/H$.
How about a case of $c_1 \neq 0$?
The physical state corresponding to
$O^{c_1}_{\msc{max}} $ includes the elements,
\begin{itemize}
\item
$\psi_0^{\al}, \ldots , \psi^{\al}_{-(\al , c_1)}$
{}~~~~if $\al \in \De_{\germ}^+$ satisfies $(\al , c_1)\geq 0$,
\item
$\chi_{\al,-1}, \ldots , \chi_{\al, (\al, c_1)+1}$
{}~~~~if $\al \in \De_{\germ}^+$ satisfies $(\al , c_1) \leq -2 $,
\item
no modes of $\chi_{\al}$, $\psi^{\al}$
{}~~~~if $\al \in \De_{\germ}^+$ satisfies $(\al , c_1) =-1 $.
\end{itemize}
These modes have the following geometrical meaning.
Let $\ca{L}_{\al}$ be the line bundle
in which $\psi^{\al}$ lives. Then  $\chi_{\al}$ is a section
of $K \otimes \ca{L}_{\al}^{~-1} $. $K \equiv T^{1,0 *}\Sigma $
is the canonical bundle of $\Sigma$.
The space of the solutions of the equation of motion
for $\psi^{\alpha}$ is $H^0 (\Sigma , \ca{L}_{\al})$,
and that for  $\chi_{\al}$ is
$H^0 (\Sigma , K \otimes \ca{L}_{\al}^{~-1})$.
These spaces can be realized by
\be
  H^0 (\bc P ^1 , \ca{L}) \cong
    \left \{
     \ba{ll}  0 ,  & \mbox{deg} \, \ca{L} < 0  \\
            P_n (X_0 , X_1 ) , & \mbox{deg} \, \ca{L} = n \geq 0 ,
     \ea  \right.
\ee
where $X_0$, $X_1$ denote the homogeneous coordinates
of $\bc P^1$, and $P_n (X_0, X_1)$ is the set of
homogeneous polynomials of order $n$.
Since we are now considering the situation
of $F(a) \sim c_1 \de^{(2)}(z- \infty)$,
we may regard as  $\ca{L} = \ca{O}(\infty)^n$,
so the $n+1$ independent elements of $H^0 (\bc P ^1 , \ca{L})$
behave as $1, z, \ldots , z^n$ around $z= 0$.
Moreover, by recalling  $\mbox{deg}\, \ca{L}_{\al}= (\al, c_1)$,
$\mbox{deg}\,  K\otimes \ca{L}_{\al}^{~-1}= -2-(\al,c_1)$
and the mode expansions
$\dsp \psi^{\al}(z) = \sum_{n\in \bz} \frac{\psi^{\al}_n}{z^n}$,
$\dsp \chi_{\al}(z) = \sum_{n\in \bz} \frac{\chi_{\al ,n}}{z^{n+1}}$,
we can find that $O_{\msc{max}}^{c_1}$ just  includes
all the solutions of the equations of motion for
$\psi^{\al}$, $\chi_{\al}$.
Therefore we can say
$O^{c_1}_{\msc{max}}$ corresponds
to the ``top cohomology class'' on the instanton moduli space.
Because of the relation;
\be
  O^{c_1 = \gamma}_{\msc{max}} = \ca{U}_{\scdminus
  \gamma} (O^{c_1=0}_{\msc{max}}) ,
\label{sflow max}
\ee
which is easily shown from \eqn{out 2}
$\ca{U}_{\scdminus \gamma}$ maps the top cohomology class on the $c_1=0$
instanton moduli space to that on the $c_1 = \gamma $ moduli space.

                      Nextly we shall study the correlator of the form
$ \langle \prod_{i=1}^N O_{a_i} \rangle_{c_1}$.
We first notice that this correlator is non-zero
if and only if the following relation holds;
\be
 \prod_{i=1}^N \, O_{a_i} \sim  \mbox{const} \, O^{c_1}_{\msc{max}}
    + \sum_b O_b
  ~(\mbox{mod BRST}).
\label{ope max}
\ee
Making use of  this observation  and combining the relations
\eqn{sflow max}, \eqn{ring sflow 2} successively, we can show
the formula;
\be
\left \langle \ca{U}_{\gamma_1} (O_{a_1})\,
  \ca{U}_{\gamma_2} (O_{a_2})\, \ldots \,
  \ca{U}_{\gamma_n} (O_{a_n}) \right\rangle_{c_1}
= \left \langle O_{a_1}\,
  O_{a_2} \, \ldots \, O_{a_n}
\right\rangle_{c_1 + \sum_{i=1}^n  \gamma_i }.
\label{main}
\ee
This  means that the  field identification rule (\ref{identify})
by the spectral
flow is still consistent  at the level of correlators
{\em if summing them up with respect to the Chern numbers $c_1$.\/}
Namely the correlation function among the identified observables
should be defined by;
\be
\left \langle \lb O_{a_1}\rb \,\lb O_{a_2}\rb \,\ldots \,
   \lb O_{a_n}\rb \right\rangle \df
\sum_{c_1 \in \ca{P}(\gerg/\gerh)}
\left \langle  O_{a_1} \, O_{a_2} \,\ldots \,
    O_{a_n} \right\rangle_{c_1} ,
\label{cor id1}
\ee
or equivalently, by fixing $c_1$ to be some definite value,
one may take the next;
\be
\left \langle \lb O_{a_1}\rb \,\lb O_{a_2}\rb \,\ldots \,
   \lb O_{a_n}\rb \right\rangle \df
\sum_{\gamma \in \ca{P}(\gerg/\gerh)}
\left \langle  \ca{U}_{\gamma }(O_{a_1}) \, O_{a_2} \,\ldots \,
    O_{a_n} \right\rangle_{c_1} .
\label{cor id2}
\ee

{}~

{}~

       It is interesting to derive the identity \eqn{main}
from somewhat different viewpoints.
We restrict ourselves to a simple case $G/H = G/T$.
We start with the definition of correlator by path-integration
;
\bea
&& \left\langle \prod_{i=1}^N O_{a_i}(z_i) \right\rangle_{c_1}
                  \nonumber \\
&&~~~~= \int \ca{D}(g,X,\chi,\psi,\zeta,\xi)\,
   \prod_{j=1}^N O_{a_j}(z_j) \, e^{-kS_G(g:\, \om) -S_X(X:\, \om)
      -S_{\chi\psi}(\chi,\psi:\, \om)-S_{\zeta\xi}(\zeta,\xi:\, \om)}.
                  \nonumber \\
&& ~~~~~~~~~
\label{path int}
\eea
$\om$ in (\ref{path int}) is the real valued back-ground gauge field;
$a=i\omega$ and
$\dsp c_1 = -\frac{1}{2\pi} \int_{\Sigma} F(\om)$.
Since we study the holomorphic part only,
we may set
$\omega^{10}=0$
by treating $\omega^{10}$ and $\omega^{01}$ as independent
variables
\footnote{Strictly speaking,
we should treat the holomorphic and the anti-holomorphic
parts simultaneously
by setting $\omega^{10 \dagger}= \omega^{01}$.}.

                           Let us consider the case of
$ \left\langle \prod_{j=1}^N O_{a_j}(z_j) \right\rangle_{c_1}$
with $\dsp c_1 = \sum_{j=1}^N \gamma_j$, $\gamma_j \in P$ and
$a_j \equiv (\hat{\La}_j, w_j)\in \hat{P}^k_+ \times W(\gerg/\gerh) $
for $\forall j$. Namely, we set all the inserted observables
semi-classical.
First  we point out that the topological invariance of our theory
implies that the correlation function (\ref{path int})
does not depend on any configuration of the back-ground
gauge field $\om$ as far as
the Chern numbers $c_1$ are fixed.
It is assured by the BRST-exactness (\ref{exactJtot})
of the total current
$J_{\msc{tot}}^{\gert}\equiv
J_g^{\gert}+ J_X^{\gert}+J^{\gert}_{\chi\psi}$.
Hence one may choose the following  configuration;
\be
  \ba{l}
\dsp \om^{01}|_{U_0} = -i\sum_{j=1}^N \gamma_j\, \deebar \log (z-z_j),  \\
\om^{01}|_{U_{\infty}} = 0 ,
\ea
\label{conf om}
\ee
where $U_0$, $U_{\infty}$ are
the coordinate patches around $0, \infty $
such that $\Sigma = U_0 \cup U_{\infty}$,
$\{z_j\} \subset U_0$, $\{ z_j \} \not\subset U_{\infty}$.
This configuration gives
\be
F(\om) = i\pi \sum_{j=1}^N \gamma_j  \, \delta^{(2)}(z-z_j) \,
                     d\bar{z}\wedge dz ,
\label{conf f}
\ee
and then $\dsp c_1 = \sum_{j=1}^N \gamma_j$.
We are now considering the situation
such that the curvature of the back-ground gauge field
has some delta function singularities
at the points $z_j$ where the observables
$O_{a_i}$ have been inserted, while
in the situation considered before
we took $\dsp F(\om) \sim \sum_j \gamma_j \, \delta^{(2)}(z-\infty)$.
These curvature singularities
will  give some non-trivial modifications
to the observables inserted at the points of singularities.

To recognize the above mentioned effect directly   it may be
helpful to consider first the $X$-sector.
Recall that the action of $X$-sector \eqn{action X} has
the term $\dsp \sim \int (X, F(\om))$.
Then the substitution of  \eqn{conf f}
into the action \eqn{action X} is  equivalent to
the insertion of some vertex operators at the points $z_j$.
This effect will add the value $-(k+g^{\vee}) \gamma_j$ on
to the $X$-momentum of $O_{a_j}(z_j)$.
It signals the possibility
that the instanton contribution modifies  the physical observables.
This story is, however, too naive to justify completely.
Because the vertex operator insertions
from the curvature singularities
are at the {\em same\/} points as that  $O_{a_j}$ are inserted,
we should treat carefully the OPE singularities among them.
We will later discuss these OPE singularities.

In order to proceed further we shall consider the following
chiral gauge transformation\footnote
   {We set $\Om^{\dag}=1$ formally,
   since we are here dealing with  the holomorphic sector only.};
\be
\Om(t)(z) = e^{tu(z)},~~~u(z) =\frac{h}{z-w},~~~ (\forall h \in \gert),
\ee
and we
replace $-kS_G (g:\, \om)$ by
$-kS_G (g: \,^{\Om(t)^{-1}}\! \om)$
in \eqn{path int}.
Then, by changing the integration variable
$g$ to $^{\Omega(t)^{-1}}g$,
we can get the identity;
\bea
\ba{l}
\dsp
\int \ca{D}(g,X,\chi,\psi,\zeta,\xi)\,
   \prod_{j=1}^N O_{a_j}(z_j) \, \exp\{-kS_G(g:\,^{\Om(t)^{-1}}\!\om) \\
\hspace{7cm}      -S_X(X:\, \om)
      -S_{\chi\psi}(\chi,\psi:\, \om)-S_{\zeta\xi}(\zeta,\xi:\, \om)\} \\
= \dsp
\int \ca{D}(g,X,\chi,\psi,\zeta,\xi)\,
   \prod_{j=1}^N e^{t\langle w_j(\La_j),u(z_j) \rangle}O_{a_j}(z_j) \,
     \exp\{-kS_G(g: \, \om) +kS_G(\Om(t):\, \om) \\
\hspace{7cm}      -S_X(X:\, \om)
      -S_{\chi\psi}(\chi,\psi:\, \om)-S_{\zeta\xi}(\zeta,\xi:\, \om)\}. \\
{}~~~~~~~~~~~~~~~~~~
\ea
\label{change}
\eea
Here we have used the fact;
$\ca{D}(\,^{\Om(t)^{-1}}\!g)=\ca{D}g$,
$O_{a_j}(z_j)\lb\,^{\Om(t)^{-1}}\!g\rb =
e^{t\langle w_j(\La_j), u(z_j)\rangle} O_{a_j}(z_j)\lb g \rb$
and the Polyakov-Wiegmann identity;
$S_G (\,^{\Om(t)^{-1}}\!g:\,^{\Om(t)^{-1}}\!\om) =
S_G(g:\, \om) - S_G (\Om(t):\,\om)$.
Differentiating the both hand sides of \eqn{change}
with respect to $t$ and then setting $t=0$,
we obtain the Ward identity;
\be
\ba{l}
\dsp \int \ca{D} (g,X, \cdots)\, (h, J^{\gert}_g(w))\prod_{j=1}^N O_{a_j}(z_j)
     \, e^{-S_{\msc{tot}}(g,X,\cdots :\, \om)} \\
\dsp ~~~~~~~= \sum_{j=1}^N
  \frac{\langle w_j(\La_j)+k\gamma_j , h \rangle}{w-z_j}
\int \ca{D}(g,X,\cdots)\, \prod_{j=1}^N O_{a_j}(z_j)
     \, e^{-S_{\msc{tot}}(g,X,\cdots :\, \om)}.
\ea
\label{insert jg}
\ee
Notice that the net effect of the curvature singularity
$\sim \gamma_j \de^{(2)}(z-z_j)$
is the shift of the weight of $g$-sector; $w_j(\La_j)~\lrarr ~
w_j(\La_j)+k\gamma_j$.
The similar arguments also work for the $X$-sector and the
$\chi\psi$-sector. By replacing $S_X(X:\, \om)$
(or $S_{\chi\psi}(\chi,\psi:\, \om)$) by $S_X(X:\,^{\Om(t)^{-1}}\!\om)$
(or $S_{\chi\psi}(\chi,\psi:\,^{\Om(t)^{-1}}\!\om)$), and
making use of the next formulas
\be
\ba{l}
 \ca{D}(\,^{\Om(t)^{-1}}\!X)=\ca{D}(X),\\
 S_X(\,^{\Om(t)^{-1}}\!X:\,^{\Om(t)^{-1}}\!\om) =
  S_X(X:\,\om)-(-k-g^{\vee})S_G(\Om(t):\,\om),\\
 \ca{D}(\,^{\Om(t)^{-1}}\!\chi,\,^{\Om(t)^{-1}}\!\psi)=
 \ca{D}(\chi,\psi)\, e^{g^{\vee}S_G(\Om(t):\,\om)},\\
 S_{\chi\psi}(\,^{\Om(t)^{-1}}\!\chi,
  \,^{\Om(t)^{-1}}\!\psi:\,^{\Om(t)^{-1}}\!\om)=
  S_{\chi\psi}(\chi,\psi:\, \om),
\ea
\ee
one can show  the Ward identities;
\be
\ba{l}
\dsp \int \ca{D} (g,X, \cdots)\, (h, J^{\gert}_X(w))\prod_{j=1}^N O_{a_j}(z_j)
     \, e^{-S_{\msc{tot}}(g,X,\cdots :\, \om)} \\
\dsp ~~~~~~~= \sum_{j=1}^N
\frac{-\langle w_j*\La_j+(k+g^{\vee})\gamma_j , h \rangle}{w-z_j}
\int \ca{D}(g,X,\cdots)\, \prod_{j=1}^N O_{a_j}(z_j)
     \, e^{-S_{\msc{tot}}(g,X,\cdots :\, \om)},
\ea
\label{insert jX}
\ee
\be
\ba{l}
\dsp \int \ca{D} (g,X, \cdots)\, (h, J^{\gert}_{\chi\psi}(w))
    \prod_{j=1}^N O_{a_j}(z_j)
     \, e^{-S_{\msc{tot}}(g,X,\cdots :\, \om)} \\
\dsp ~~~~~~~= \sum_{j=1}^N \frac{\langle w_j(\rho_{\gerg})-\rho_{\gerg}
               +g^{\vee}\gamma_j , h \rangle}{w-z_j}
\int \ca{D}(g,X,\cdots)\, \prod_{j=1}^N O_{a_j}(z_j)
     \, e^{-S_{\msc{tot}}(g,X,\cdots :\, \om)}.
\ea
\label{insert j chi psi}
\ee

                 What do the identities \eqn{insert jg}, \eqn{insert jX},
\eqn{insert j chi psi} mean?
These identities suggest that,
\begin{em}
under the singular back-ground
$F(\om)\sim \gamma_j \de^{(2)}(z-z_j)$,
 the chiral primary field
$O_{a_j}(z_j)$ behaves as $\ca{U}_{\gamma_j}(O_{a_j})(z_j)$
in the flat back-ground.
\end{em}
Because the BRST-invariance is not broken even in the non-flat
back-ground, the BRST-invariant operator $O_{a_j}(z_j)$
should be changed  to another BRST-invariant operator by the effect of
curvature singularity.
Comparing the spectrum of physical observables studied in the previous
section with the data of the Ward identities,
we can conclude that $O_{a_j}(z_j)$ should  be
changed to the operator $\ca{U}_{\gamma_j}(O_{a_j})(z_j)$.
So, we have again obtained the relation \eqn{main}.
Notice that this derivation of \eqn{main}
does not need the result \eqn{ring sflow 2}
which played an important role to obtain the relation
\eqn{main} in the previous section.
We can rather show this formula \eqn{ring sflow 2}
very easily by means of the above result.
In fact, if noting that
\be
\left \langle \ca{U}_{\gamma_1} (O_{a_1})\,
  \ca{U}_{\gamma_2} (O_{a_2})\,
\ca{U}_{\scdminus (\gamma_1\scdplus\gamma_2)}(O_{a_3}) \,
\prod_{\msc{others}}O_b \,
\right\rangle_{c_1}
= \left \langle O_{a_1}\,
  O_{a_2} \, O_{a_3}\, \prod_{\msc{others}}O_b \,
\right\rangle_{c_1},
\ee
which is a trivial identity derived from \eqn{main},
the desired result \eqn{ring sflow 2} follows.
Thus the relation \eqn{ring sflow 2} is a
natural consequence of the topological
invariance of our model.

                  Related with the above discussion based
on the path integration
it may be helpful to
study the local operator formulation around the point
where the back-ground gauge field is singular.
Consider again \eqn{conf om} as the configuration of the back-ground
gauge field and
pay attention to  a particular point, say,  $z_1$.
We take a holomorphic coordinate $\{ z,U\}$ around $z_1$ such that
$z_1 = 0$, $z_j \not\in U$ $(j\neq 1)$
and simply write as $\gamma_1 = \gamma$, namely,
\be
 \om^{01}|_{U} = -i \gamma\, \deebar \log z .
\label{bg}
\ee
Let us consider the operator formalism on $U$.
Notice that,
by setting
$\Om_{(\gamma)}= z^{-\nu^{-1}(\gamma)}$,
$\om^{01}$
can be trivialized on $U$;
\be
i\om^{01}|_U =
 -\deebar \Om_{(\gamma)}\,\Om_{(\gamma)}^{~-1}|_U .
\label{bg 2}
\ee
Thanks to this relation we can consistently give the operators
appropriate to the back-ground \eqn{bg}.
Let $\ca{O}=\ca{O}\lb g, X, \chi,\psi \rb$ be a general operator
under the flat back-ground,
then the corresponding operator $\ca{O}^{(\gamma)}$
which is appropriate
to the singular back-ground \eqn{bg} (or \eqn{bg 2})
should be defined
by the gauge transform
$\ca{O}^{(\gamma)} =\,^{\Om_{(\gamma)}}\!\ca{O}$, that is,
\be
\ba{lll}
 \ca{O}^{(\gamma)} \lb g,X,\chi,\psi \rb &\df& \ca{O}
  \lb \,^{\Om_{(\gamma)}^{~-1}}\!g, \,^{\Om_{(\gamma)}^{~-1}}\!X,
  \,^{\Om_{(\gamma)}^{~-1}}\!\chi, \,^{\Om_{(\gamma)}^{~-1}}\!\psi \rb \\
   & \equiv  & \ca{O}
   \lb \Om_{(\gamma)}^{~-1}g,\, X+\al_+ \log \Om_{(\gamma)}^{~-1},\,
     \Om_{(\gamma)}^{~-1}\chi \Om_{(\gamma)} ,\,
     \Om_{(\gamma)}^{~-1}\psi \Om_{(\gamma)}\rb.
\ea
\ee
In particular the current $J_g$  is transformed to
$J_g^{(\gamma)}$;
\be
\ba{lll}
  J_g^{(\gamma)}(z) & = & J_g (z)\lb \Om_{(\gamma)}^{~-1}g \rb \\
   & = & \Om_{(\gamma)}^{~-1}\, J_g(z)\, \Om_{(\gamma)}
    + k \Om_{(\gamma)}^{~-1}\, \partial  \Om_{(\gamma)} (z) .
\ea
\label{jg gamma}
\ee
If we substitute $\Om_{(\gamma)}(z) = z^{-\nu^{-1}(\gamma)}$
into \eqn{jg gamma} and compare it with the defintion
of spectral flows \eqn{sflow g},
we can find the following identity;
\be
  J_g^{(\gamma)}(z) = \ca{U}_{\gamma}\, J_g(z) \, \ca{U}_{\gamma}^{~-1}.
\label{sflow gauge tr}
\ee
It is easy to observe that the same relations also hold for other fields;
\be
\ba{l}
X^{(\gamma)}(z) = \ca{U}_{\gamma}\, X(z) \, \ca{U}_{\gamma}^{~-1}, \\
\chi^{(\gamma)}(z) = \ca{U}_{\gamma}\, \chi(z) \, \ca{U}_{\gamma}^{~-1}, ~~~
\psi^{(\gamma)}(z) = \ca{U}_{\gamma}\, \psi(z) \, \ca{U}_{\gamma}^{~-1}.
\ea
\label{sflow gauge tr 2}
\ee
These mean that the singular gauge transformation $\Omega_{(\gamma)}$
which connects the flat back-ground (on $U$) to the singular back-ground
\eqn{bg} is realized by the spectral flow $\ca{U}_{\gamma}$.
These transformed operators will characterize
the state vector which is inserted at the origin of the
coordinate patch $U$.
Especially the vacuum vector under the singular back-ground
(\ref{bg}) will be given by
\be
|0\rangle_{(\gamma)} = \ca{U}_{\gamma}|0\rangle,
\label{vacuum gamma}
\ee
where $|0\rangle$ is the usual vacuum with the
locally flat back-ground on $U$, that is,
$\tilde{\omega}$ such that
$\tilde{\omega}=^{\Omega_{(\gamma)}^{-1}}\omega $.
The gauge transformed
physical observables can be written as
\be
O_a^{(\gamma)}(z) = \ca{U}_{\gamma}\, O_a(z)\, \ca{U}_{\gamma}^{~-1}.
\label{oa gamma}
\ee
Hence the next readily follows;
\be
 O_a^{(\gamma)}(0)|0\rangle_{(\gamma)}
=\ca
{U}_{\gamma}(O_a)(0) | 0\rangle.
\label{sflow relation}
\ee
This relation can be read as follows:
In the L.H.S of \eqn{sflow relation} the ``topological charge''
$\gamma$ is attached to the vacuum vector $|0\rangle_{(\gamma)}$,
which correspond to the existence of curvature singularity.
While, in the RHS of \eqn{sflow relation} the vacuum vector
has no charge and  the charge $\gamma$ is absorbed into the operator
$O_a$, which means $O_a$ is changed to  $\ca{U}_{\gamma}(O_a)$.

                                For a general operator $\ca{O}(z)$
its gauge transformed partner $\ca{O}^{(\gamma)}(z)$
will have a singularity at $z=0$.
This is because, in general,
$\ca{O}^{(\gamma)}$ has the same singularity as that
of $\Omega_{(\gamma)}$.
Can we define the operator $\ca{O}^{(\gamma)}(0)$
in the L.H.S of (\ref{sflow relation})?
This issue has its origin in the singular configuration
of the back-ground gauge field (\ref{bg}) and
the same difficulity has appeared in the earlier discussion
of the $X$-sector,
where it is necessary to insert some vertex operators at the points
other operators have been inserted.
We cannot avoid this difficulity if we study each sector
independently.
However, fortunately we can find,
from the defintions of spectral flow \eqn{sflow g}, ...,
that the singularity of each sector precisely cancells in all.
This is not surprising, since the BRST-invariance
(in particular, the $Q_{T^{\bc}}$-invariance) of the  operator $O_a$
leads to the chiral gauge invariance.
Actually $O_a^{(\gamma)}(z)$ should define the same cohomology
class as that of $ O_a(z)$, namely,
\be
  O_a^{(\gamma)}(z) = O_a (z)  + \{ Q_{\msc{tot}}, * \}
\ee
should hold.
So we may rewrite \eqn{sflow relation}  as
\be
 O_a(0)|0\rangle_{(\gamma)}
    =\ca{U}_{\gamma}(O_a)(0) | 0\rangle
    + Q_{\msc{tot}}| * \rangle .
\label{sflow relation 2}
\ee
To sum up, one may say:
When (and only when) treating the $g$, $X$, $\chi\psi$-sectors
{\em as a combined system\/} and
considering  only the BRST-invariant oprators,
one can construct a consistent operator
formalism even under the singular back-ground \eqn{bg}.
This reflects the fact that the total system is  topologically invariant
but each sector is not.

                     It may be also interesting to consider
the following configuration
of the back-ground gauge field instead of \eqn{bg};
\be
\om|_U = -\gamma \, \om(\bm{g})|_U,
\label{bg 3}
\ee
where $\om(\bm{g})$ is the Levi-Civita connection.
Since the Euler number of  $U \cong \mbox{\em{semisphere,}}$
is equal to 1, $\dsp -\frac{1}{2\pi}\int_U F(\om)= \gamma$ holds.
Under this configuration of $\omega$ (\ref{bg 3}) the conformal
fields in our model will gain some extra spins.
For example, the ghost $\psi^{\al}(z)$ will be transformed to
\be
\ba{l}
\dsp \psi^{(\gamma)\al}(z)(\equiv \ca{U}_{\gamma}\, \psi^{\al}(z)\,
\ca{U}_{\gamma}^{~-1})    =
\sum_{n\in \bz} \frac{\psi^{\al}_n}{z^{n-(\al,\gamma)}},   \\
\dsp ~~~~~~~~~~ =
\sum_{n\in \bz} \frac{\psi^{(\gamma)\al}_n}{z^{n}},
\ea
\label{psi al gamma}
\ee
where
\be
\psi^{(\gamma)\al}_n (\equiv \ca{U}_{\gamma}\, \psi^{\al}_n \,
    \ca{U}_{\gamma}^{~-1}) = \psi^{\al}_{n+(\al,\gamma)} .
\ee
The first line of \eqn{psi al gamma} and the vacuum condition;
$\psi_n^{\al}|0\rangle_{(\gamma)} = 0$ iff $n> (\al,\gamma)$,
 mean that the field $\psi^{(\gamma)\al}(z)$
behaves as a spin $-(\al,\gamma)$ primary field
on the vacuum $|0\rangle_{(\gamma)}$
with respect to the modes $\psi^{\al}_n$.
This aspect  corresponds to the configuration \eqn{bg 3}.
one may say that the spectral flow is
the transformation changing the spins of the elementary fields
as is expected from the back-ground \eqn{bg 3}.
On the other hand, the 2nd line of \eqn{psi al gamma}
and the vacuum condition; $\psi^{(\gamma)\al}_n |0\rangle_{(\gamma)}=0$
iff $n>0$, say that  it is a spin 0 primary field
with respect to the gauge transformed modes $\psi^{(\gamma)\al}_n$,
which  corresponds to the configuration \eqn{bg} considered before.
In this way,  we obtain the  two physical interpretations
of the spectral flow corresponding to the choice of the back-ground
\eqn{bg} or \eqn{bg 3}; one of them is
a singular gauge transformation and the other is a spin changing
transformation.
They are of course consistent with each other.

{}~

\section{Conclusions}
\cleqn

{}~

We have investigated the topological gauged WZW models for
the cases of general K\"{a}hler homogeneous spaces.
The gauge fixing and the operator formalism based on it
were presented in such a way that they are natural extensions of  those
in \cite{ns}.
We investigated   the BRST-cohomology of our system,
defined by the total BRST-charge
$Q_{\msc{tot}}= Q_{G/H}+ Q_{Z(H^{\bc})}+ Q_{H_0^{\bc}}$.
The cornerstone of this study is the concept ``spectral flows''
\cite{LVW,HT}, which  are symmetry  transformations  preserving
the BRST-charge $Q_{\msc{tot}}$.
Because of the relation
\eqn{ring sflow 2};
$\ca{U}_{\gamma_1}(O_a)
\ca{U}_{\gamma_2}(O_b)=\sum_{c \in I}
C_{a~b}^{c}\ca{U}_{\gamma_1 \scdplus \gamma_2}(O_{c})$,
the identifications of the physical obervables
by the spectral flows are compatible with the chiral ring structure.
We further argued the problem of these field identifications from the
global geometrical standpoint.
The correlation function among the identified observables,
which is valid under any back-ground gauge field,
is described in \eqn{cor id1}(or \eqn{cor id2}).

                   Under these studies the geometrical and physical
pictures of the spectral flows appeared.
They give the following two insights
on the roles of the gauge field.

                   Firstly, the spectral flow is capable to connect
two arbitary instanton sectors in the system which are labelled by
different Chern numbers.
This is recoginized as a phenomenon that the physical observables
are transmuted by the effect of the back-ground curvature singularity,
and it gives the relation \eqn{main}.
Moreover, the consistency condition \eqn{ring sflow}
of the ring structure under the field identifications \eqn{identify}
was shown to be a natural consequence of the relation \eqn{main}.
In this respect it may be important to remark the analogy with
the two-dimensional $BF$ gauge theory \cite{BT}:
$\dsp S_{\msc{BF}}=\frac{i}{2 \pi}\int (\phi,F(\omega))$.
In this $BF$ theory the correlator among the vertex operators
$e^{i(\beta,\phi)}$ satisfies the same relation as in \eqn{main}.
The above $BF$ theoretical aspect  of topological
$SU(2)/U(1)$ gauged WZW model played an important part in the
study of two-dimensional topological gravity \cite{witten}.

                    Secondly, related with the path-integral approach,
it was shown that two physical interpretations were possible
for the spectral flow.
It can be interpreted as a transformation changing the spins of
the elementary  fields in the system, while
it can also work as a singular gauge transformation which
creates an appropriate back-ground charge on the physical vacuum.
In this respect it is useful to note the analogy
with the following Coulomb gas system:
$\dsp S_{\msc{CG}=\frac{1}{4 \pi i}}\int \{(\bar{\partial}\phi,\partial \phi)
+2\alpha_+(\phi,F(\omega))+2\alpha_-\rho_{\gerg}(\phi)R(\bm{g}) \}$.
If we set $\omega=-\omega(\bm{g})\rho_{\gerg}$,
the EM tensor of the system $T_{\msc{CG}}$ will suffer
the twist by the $U(1)$-current $J_{\msc{CG}}=2i\alpha_+\partial \phi$;
$T_{\msc{CG}}\rightarrow \tilde{T}_{\msc{CG}}
=T_{\msc{CG}}+\frac{1}{2}\partial J_{\msc{CG}}$.
This causes the changes of the spins of the vertex operators
$e^{i(\beta,\phi)}$.
The important difference between this Coulomb gas system and
our topological system is that, in the topological model,
the procedure of changing the spins is performed BRST-invariantly,
because both the EM tensor and the $U(1)$-current are BRST-exact.
The above "twist" does not change the physical
contents of the topological system,
while it does in the Coulomb gas system.

{}~

                      In order that spectral flows work completely
in a given model,
the underlying current algebras must be integrable.
(See appendix B.)
Therefore, if
they are non-integrable,
the symmetry of spectral flow
may  be lost.  In these cases the state identification
(in the sence of  this paper) will no longer work,
and so the {\em infinite\/} BRST  states will be left for us.
It seems interesting to understand these infinite physical states
in a unifying standpoint, for example, by some appropriate
generalization of spectral flow.
In \cite{ns2} we plan to study  the case of  $SU(2)/U(1)$
with fractional levels.
Especially the infinte physical states in the system
will be studied from the above point of view.
The correspondence between this infinite dimensional cohomology
and the Lian-Zuckerman cohomology \cite{LZ} of Liouville theory
in two-dimensional gravity will be discussed.

{}~

\section*{Acknowledgements}
We thank Mr.Y.Sugiyama for the discussion at the early stage of this work.


\newpage

\appendix
{\Large \bf Appendix}
\section{Quantization of the $N=2$ SUSY Gauged WZW Model}
\cleqn

{}~

                          In this appendix we present the quantizations  of
$N=2$ supersymmetric (SUSY) gauged WZW models
\cite{witten,Nakatsu}
in order to comment the relation between
the "twist" of the SUSY gauged WZW models
and that of $N=2$ SCFTs \cite{EY}.
We note that
these quantizations are
essentially
a special case of the work in \cite{Schnitzer},
in which general (not necesarily of $N=2$) SUSY gauged WZW models
were considered and it was
shown that they describe  $N=1$ supercoset CFTs.

                     $N=2$ SUSY gauged WZW model associated with $G/H$,
where $G/H$ is assumed to be K\"{a}hler, is given by
\be
Z= \int \ca{D}g\ca{D}A\ca{D}\bm{\psi}\ca{D}{\bar{\bm{\psi}}}\,
  \exp \left \lb  - kS_G(g,A) - \frac{1}{ \pi }
\int_{\Sigma} dv(\bm{g}) \,
  \{ (\bm{\psi}, \deebar_{A \bar{z} }\bm{\psi}  )+
(\bar{\bm{\psi}}, \partial_{Az}\bar{\bm{\psi}})\}\right\rb.
\label{susy gwzw}
\ee
In this expression
the Weyl fermions
$\bm{\psi}$, $\bar{\bm{\psi}}$
are $\germ_+ \oplus \germ_- $-valued,
where we set $\germ_{\pm}$ be  those
introduced in section 2 in the text.
The gauge field $A$ is $\gerh$-valued.
$dv(\bm{g})$ denotes   the canonical volume element defined by
a fixed K\"{a}hler metric $\bm{g}$.

As in the topological case we must perform the gauge fixing
for the $H^{\bc}$-chiral gauge transformations,
and for this aim we need to estimate the chiral anomalies
of the ingredients.
The solution of this problem is almost the same as the topological case
in the text.
The only exception is the abelian part (i.e. $Z(H^{\bc})$-part) of the
anomaly of the coset fermions $\bm{\psi}$, $\bar{\bm{\psi}}$.
It is due to the fact that $\bm{\psi}$, $\bar{\bm{\psi}}$ are spinor
fields and that we have to use the index theorem for the spin complex
rather than the Dolbeault complex.
Therefore we must replace the estimation of the anomaly of
$\chi \psi$-system \eqn{determinant chi psi} by the following one;
\bea
Z_{\bm{\psi}} &= &
    \int \ca{D}(\bm{\psi}, \bar{\bm{\psi}})\,
    \exp \left\lb -\frac{1}{\pi }
      \int_{\Sigma} dv(\bm{g}) \,
  \{ (\bm{\psi}, \deebar_{{}^h\!a \bar{z} }\bm{\psi}  )+
(\bar{\bm{\psi}}, \partial_{{}^h\!a z}\bar{\bm{\psi}})\}\right\rb  \nonumber\\
     &=& \int \ca{D}(\bm{\psi}, \bar{\bm{\psi}})\,
    \exp \left\lb -\frac{1}{\pi}
      \int_{\Sigma} dv(\bm{g}) \,
  \{ (\bm{\psi}, \deebar_{a \bar{z} }\bm{\psi}  )+
(\bar{\bm{\psi}}, \partial_{az}\bar{\bm{\psi}})\}\right\rb \nonumber\\
       & &\dsp  ~~~\times \prod_{i=1}^r
   \exp \left \{(g^{\vee}-h_i^{\vee}) S_{H_0^{(i)}}(\rho^{(i)} , a_0^{(i)})
       \right\} \nonumber \\
     & & ~~~\times \exp \left\lb
     \frac{i g^{\vee}}{4 \pi} \int_{\Sigma} \{
      (\deebar X,\partial X)
     +2i (X, F(\om))  \} \right\rb .
\label{determinant coset fermion}
\eea
Notice that
a distinction from \eqn{determinant chi psi} is
the absence of the ``back-ground charge term''
$\dsp  \sim \int \rho_{\gerg/\gerh}(X)R$.
In this way we can finally arrive at the following
expression of the gauge-fixed system ;
\bea
 Z_{\msc{g.f.}}\lb \bm{g} \rb & = & \int \ca{D}a Z_{\msc{g.f.}}
                                                  \lb a, \bm{g} \rb \equiv
     \int \ca{D}(a_0, \om)
       Z_{\msc{g.f.}}\lb a_0, \om, \bm{g} \rb,  \nonumber \\
Z_{\msc{g.f.}} \lb a_0, \om, \bm{g} \rb &=&
\int \ca{D}(g, \rho, X, \bm{\psi}, \bar{\bm{\psi}},
   \zeta, \bar{\zeta}, \xi, \bar{\xi} )  \nonumber \\
  &  &  ~~\times \exp \{
       -k S_G ( g, a)  -
     S_{\bm{\psi}} (\bm{\psi} ,\bar{\bm{\psi}}, a ) \} \nonumber \\
&  & \dsp  ~~\times \exp \left\{
   \sum_{i=1}^r (k + g^{\vee}+ h_i^{\vee}) S_{H_0^{(i)}}
          (\rho^{(i)}, a_0^{(i)})
             \right.  \nonumber  \\
& & \dsp  \left.   ~~~~~~~~~~~~- S_X' (X , \om, \om(\bm{g}))
      - S_{\zeta \xi} (\zeta, \bar{\zeta}, \xi, \bar{\xi} , a_0) \right\},
\label{gauge fixed app}
\eea
where we set
\bea
S_{\bm{\psi}} (\bm{\psi},\bar{\bm{\psi}}, a )
      & = & \frac{1}{\pi } \int_{\Sigma} dv(\bm{g}) \,
  \{ (\bm{\psi}, \deebar_{a \bar{z} }\bm{\psi}  )+
(\bar{\bm{\psi}}, \partial_{az}\bar{\bm{\psi}})\}
                            \label{action  psi}\\
S_{\zeta \xi} (\zeta, \bar{\zeta}, \xi, \bar{\xi} , a_0)
& = & \frac{1}{2 \pi i} \int \{ (\deebar _{a_0} \xi, \zeta)
-(\bar{\zeta},\partial_{a_0} \bar{\xi})   \}
\label{action zeta xi app}\\
S'_X (X , \om, \om(\bm{g}) )
& = & \frac{ 1}{4 \pi i} \int_{\Sigma} \left\{ (\deebar X,\partial X)
     +   2i \al_+ (X, F(\om)) \right\}   \nonumber \\
   & &   \label{action X app}\\
   & & ~~~~~~~(\al_+ = \sqrt{k+ g^{\vee}},~
\al_-= -\frac{1}{\sqrt{k+g^{\vee}}})  . \nonumber
\eea
The fields $X$, $\rho$, $\xi$, $\bar{\xi}$, $\zeta$,
$\bar{\zeta}$ are the counterparts of those in
\eqn{gauge fixed}.

                 We can read the total EM tensor of the gauge fixed system
from \eqn{gauge fixed app};
\be
T'_{\msc{tot}} = T_g + T_{\rho} + T'_X + T_{\bm{\psi}}+ T_{\zeta \xi},
\label{ttot app}
\ee
where $T_g$, $T_{\rho}$, $T_{\zeta \xi}$ are
those given in \eqn{tg}, \eqn{trho}, \eqn{t zeta xi} and
\bea
T'_X & = & - \frac{1}{2} :(\partial_z X, \partial_z X): ,
                           \label{tx app} \\
T_{\bm{\psi}} & = & - \frac{1}{2} :(\bm{\psi}, \partial_z \bm{\psi}): .
                        \label{tpsi app}
\eea
The total central charge is easily calculated;
\bea
c'_{\msc{tot}} &=& c_g +c_{\rho}+ c'_X + c_{\bm{\psi}} + c_{\zeta \xi}
\nonumber\\
  & = & \frac{k \dimn{\gerg}}{k+g^{\vee}} +
  \sum_{i=1}^r \frac{-(k+g^{\vee}+h_i^{\vee})
  \dimn \gerh^{(i)}_0}{-(k+g^{\vee}+h_i^{\vee})+ h_i^{\vee}}
      + l    \nonumber \\
   & & ~~~~~~~~~~+  \frac{1}{2} \dimn \, (\germ_+  \oplus \germ_- )
   + (-2)\times (\dimn \gerh_0 +l)   \nonumber \\
   & =& 3 \dimn \germ_+ - \frac{12}{k+g^{\vee}} \rho^2_{\gerg/\gerh}
   \label{cks} .
\eea
It is equal to the central charge of the Kazama-Suzuki model for $G/H$.
In fact the total system is equivalent to the Kazama-Suzuki model modulo
a BRST exact term.
$T'_{\msc{tot}}$ can be factorized to
\bea
  T'_{\msc{tot}}  = T_{\msc{KS}} + T'_{Z(H^{\bc})} + T_{H_0^{\bc}},
\eea
where $T_{\msc{KS}}$ is the EM tensor of the Kazama-Suzuki model \cite{KaS}
\bea
T_{\msc{KS}} = \frac{1}{2(k+g^{\vee})} \left
    \{ \nprod (J_g, J_g)\nprod  -
 \nprod (\hat{J}^{\gerh},\hat{J}^{\gerh} )\nprod  \right \}
    - \frac{1}{2} :(\bm{\psi}, \partial_z \bm{\psi} ):, \label{tks}
\eea
and $T'_{Z(H^{\bc})}$ is defined by
\bea
T'_{Z(H^{\bc})} =
\frac{1}{2(k+g^{\vee})} \nprod (\hat{J}^{Z(\gerh)}, \hat{J}^{Z(\gerh)})\nprod
   - \frac{1}{2}:(\partial_z X, \partial_z X)  : .
\eea
$T_{H_0^{\bc}}$ is that given in \eqn{th0c}.
The physical degrees of freedom in this SUSY model are characterized by
the BRST-charge
$Q'_{\msc{tot}}=Q_{Z(H^{\bc})}+Q_{H_0^{\bc}}$
($Q_{Z(H^{\bc})}$ ,$Q_{H_0^{\bc}}$ \eqn{qhc}).
It is easy to see
\bea
  T'_{\msc{tot}} =
 T_{\msc{KS}} + \{Q'_{\msc{tot}}
    , G^{-'}_{Z(H^{\bc})} + G^-_{H_0^{\bc}} \}.
\eea
Here $G^-_{H_0^{\bc}}$ is that given in \eqn{G-HO},
and $G^{-'}_{Z(H^{\bc})}$ is
\bea
G^{-'}_{Z(H^{\bc})}  =
      -\frac{\al_-}{\sqrt{2}}
      (\zeta_z , \hat{J}^{Z(\gerh)}-  J_X ).
                   \label{g-hc app}
\eea
In this way we can see that
$\{ G^{\pm}_{G/H}$
\eqn{g+gh} \eqn{G-GH}, $T_{\msc{KS}}$, $J_{G/H}$ \eqn{jgh} $\}$
 are non-trivial physical observables
(i.e. $Q'_{\msc{tot}}$-invariant and not exact).
They generates a $N=2$ SCA.
This means that the model
is equivalent to the Kazama-Suzuki model.

             By comparing $T'_{\msc{tot}}$ with $T_{\msc{tot}}$ \eqn{Ttot2}
we can find
\bea
T_{\msc{tot}}-T'_{\msc{tot}} &=& T_{G/H} + T_{Z(H^{\bc})}
   - T_{\msc{KS}} -T_{Z(H^{\bc})}' \nonumber \\
  & = & \frac{1}{2} \partial_z J_{\msc{G/H}} + \{ Q_{Z(H^{\bc})}, * \}.
\eea
Thus the "twist" of the EM tensor of this SUSY gauged WZW model
is given by adding the term,
$$ \dsp \frac{1}{2} \partial_z J_{\msc{G/H}}\equiv
  \frac{1}{4} \partial_z :(\bm{\psi} , \bm{\psi}): +
\frac{1}{k+g^{\vee}} \, \partial_z
\rho_{\gerg/\gerh}(\hat{J}^{Z(\gerh)}). $$
The first  component  reflects  the difference of spin between
$\bm{\psi}$ and $\chi \psi$ and the second one
is due to the distinction of the abelian anomaly commented above.

\newpage

\section{Notes on Algebra Automorphisms of Affine Lie Algebras
and the Spectral Flow}
\cleqn

{}~

                      The purpose of this appendix is
to give the precise definitions of
the spectral flow \cite{LVW,HT} which is introduced  in section 4.
We will present several results needed
for our main subjects without proof.
Refer the ref.\cite{HT,Kac,KW} for the proofs and
more complete discussions.

           Let us start with preparing some notations
and conventions of affine Lie algebras.

\subsection{Notations for Affine Lie Algebras}

{}~

                  Let $\gerg$ be a (complex) simple Lie algebra
(rank $l$), and $\hat{\gerg} \equiv L \gerg \oplus \bc K \oplus \bc d$
be the corresponding affine Lie algebra.
$L \gerg \equiv \gerg \otimes \bc \lb z, z^{-1} \rb$ is
the loop algebra of $\gerg$. $K$ is the canonical central element
and $d (\dsp \equiv z \frac{d}{dz})$ is the scaling element.
We assume that $\gerg$ is simply-laced for simplicity.
Let $\gert$, $\hat{\gert}\equiv \gert \oplus \bc K \oplus \bc d$ be
the Cartan subalgebras of $\gerg$, $\hat{\gerg}$ respectively,
$\Delta = \Delta ^+ \coprod \Delta^-$,
$\hat{\Delta} = \hat{\Delta} ^+ \coprod \hat{\Delta}^-$
be the root systems of $\gerg$, $\hat{\gerg}$ and
$\Pi = \{ \al_1 , \ldots , \al_l \}$,
$\hat{\Pi} = \{ \al_0 \equiv \delta - \theta , \al_1, \ldots , \al_l \}$
($\theta$ is the highest root of $\gerg$ and $\delta$
is defined by $\delta(d) = 1 $,  $ \delta (\gert) = \delta(K)=0$ )
be the simple roots of $\gerg$, $\hat{\gerg}$.

               We denote by $(~,~)$ the Cartan-Killing metric normalized
so that the square length of each root is $2$.
It is well-known that this metric is naturally extended to
an invariant metric on $\hat{\gerg}$ \cite{Kac},
which we shall denote by $(~|~)$;
\be
\ba{l}
 \dsp  (u(z) | v(z) ) =  \frac{1}{2\pi i} \oint \frac{1}{z}
                     (u(z),v(z))\, dz ~~~(^{\forall}
 u(z) ,v(z) \in L\gerg),  \\
  (K|d ) =1,   \\
  (K| u(z)) = (d | u(z)) = (K|K) =(d|d) =0   ~~~(^{\forall} u(z)\in L\gerg) .
\ea
\ee
This metric induces the dual metric on $\hat{\gerg}^*$,
especially on $\hat{\gert}^* \equiv \gert^* \oplus
\bc \La_0 \oplus \bc \delta$,
which we express by the same notation $(~|~)$;
\be
\ba{l}
 (\al | \beta ) = (\al , \beta ) ~~~(^{\forall} \al , \beta \in \gert^*), \\
(\La_0 | \delta) = 1 ,\\
(\al |\La_0) = (\al | \delta) = (\La_0| \La_0) = (\delta|\delta) =0 ,
   ~~~(^{\forall} \al \in \gert^*) .
\ea
\ee

                         The sets of real and imaginary roots of
$\hat{\gerg}$ are denoted by
$\hat{\Delta}_{\msc{real}}
=\hat{\Delta}^{+}_{\msc{real}}\coprod \hat{\Delta}^{-}_{\msc{real}}$,
$\hat{\Delta}_{\msc{im}}
=\hat{\Delta}^{+}_{\msc{im}}\coprod \hat{\Delta}^{-}_{\msc{im}}$
respectively;
\be
\ba{l}
  \hat{\De}^+_{\msc{real}} = \Delta^+ \cup
  \{ ~ \alpha +n \delta ~;~\alpha \in \Delta,~n \in \bz_{>0}~\}, \\
  \hat{\De}^-_{\msc{real}} = -\hat{\De}^+_{\msc{real}}, \\
  \hat{\De}^+_{\msc{im}} = \{ ~ n\delta~ ;~ n\in \bz_{>0}~ \}, ~~~
  \hat{\De}^-_{\msc{im}} = - \hat{\De}^+_{\msc{im}},
\ea
\ee
where $\Delta(\Delta^+)$ are the sets of (positive) roots of $\gerg$.

                        We introduce the root lattice $\dsp Q$
and the weight lattice $\dsp P$
of $\gerg$;
\be
\dsp Q = \sum_{i=1}^l \bz \al_i,~~
\dsp P = \sum_{i=1}^l \bz \La_i,
\ee
where $\La_1, \cdots ,\La_l$ are the fundamental weights of $\gerg$
which satisfy,
\bea
 (\La_i | \al_j) = \delta_{ij}  ,
 ~~~(\La_i | \La_0)=(\La_i |\delta)=0 , \nonumber
\eea
for $1\leq ^{\forall} i,j \leq l.$
The set of dominant integral weights with level $k$ of $\hat{\gerg}$
is given by
\be
\hat{P}_+^{k}=
\{ \hat{\La}=\La +k \La_0~:~\La \in P_+,~(\La |\theta) \leq k~\},
\ee
where $P_+$ is the set of dominant integral weights of $\gerg$.

                        Let $W, \hat{W}$ be the Weyl groups of
$\gerg, \hat{\gerg}$.
$\hat{W}$ has the structure $\hat{W}=W \sdprod Q$, where the
root lattice $Q$ acts on $\hat{\gert}^*$ by "translations";
\bea
 t_{\gamma}(\hat{\mu})&=& \hat{\mu} +
 (\hat{\mu}|\delta ) \gamma -
 \left(  (\hat{\mu} | \gamma) +
 \frac{1}{2} |\gamma|^2
 (\hat{\mu}|\delta) \right) \delta
 \label{translation1} \\
 & & \hspace{3cm}
 (^{\forall} \gamma \in Q,~ ^{\forall} \hat{\mu} \in \hat{\gert}^*)
           \nonumber
\eea
Under the linear isomorphism
$\nu~ :~\hat{\gert}~  \mapru{\cong} ~ \hat{\gert^*}$
defined by $\langle \nu (h) , h' \rangle  = ( h | h')$,
\footnote{$\langle ~,~ \rangle$ is the dual pairing,$^{\mbox{i.e}}$
$\langle \nu (h),h' \rangle =\nu (h)(h').$
Especially $\nu (K)= \delta,\nu (d)= \La_0.$}
the root lattice $Q$ also acts on $\hat{\gert}$ by
\bea
 t_{\gamma} (\hat{h}) & =&  \hat{h} +
 (\hat{h}|K)\nu^{-1}(\gamma) -
 \left(  (\hat{h}|\nu^{-1}(\gamma)) +
 \frac{1}{2} |\gamma|^2 (\hat{h}|K) \right) K
 \label{translation2} \\
& & \hspace{3cm} (^{\forall} \gamma \in Q,~ ^{\forall}
                         \hat{h}\in \hat{\gert} ).  \nonumber
\eea
The Weyl group $W$ of $\gerg$ is generated by the
"reflections in simple roots" $r_{\al_{i}}(1 \leq i \leq l)$ which
act on $\hat{\la} \in \hat{\gert}^*$ by
\bea
r_{\al_i}(\hat{\la})=\hat{\la}-
(\hat{\la}|\al_i)\al_{i}~~~~(1 \leq i \leq l). \nonumber
\eea
The Weyl group $W$ may be realized as $N(T)/T$,
where $T$ is the Cartan torus of $G$ and $N(T)$ is the normalizer
of $T$.
$r_{\al_i} \in W (1 \leq i \leq l)$ will correspond to
$e^{\frac{\pi i}{2}(e_{\al_i}+e^{\al_{i}})} \in N(T)$,
where $e_{\al} \in \gerg_{\al}$ and $e^{\al} \in \gerg_{-\al}
(\al \in \Delta^+)$ are the Cartan-Weyl base of $\gerg$.
Hence $r_{\al_i}\in W (1 \leq i \leq l)$ will act on
$ \hat{x} \in \hat{\gerg}$
by
\bea
r_{\al_i}(\hat{x})=
e^{\frac{\pi i}{2}(e_{\al_i}+e^{\al_{i}})} \hat{x}
e^{-\frac{\pi i}{2}(e_{\al_i}+e^{\al_{i}})}.  \nonumber
\label{cwtg}
\eea

{}~

\subsection{Some Automorphisms of Affine Lie Algebras}

{}~

             According to \cite{HT,KW},
we shall extend the affine Weyl group $\hat{W}$
to the following group;
\be
\tilde{W} = W \sdprod P ~ (\supset \hat{W}) ,
\label{tilde W}
\ee
where the classical weight lattice $P$ acts on $\hat{\gert}^*$
$(\hat{\gert})$ in the completely same way as the root lattice $Q$
(see \eqn{translation1}, \eqn{translation2}.)
Moreover we  introduce the following subgroup of $\tilde{W}$;
\be
D = \{ \hat{w} \in \tilde{W} ~;~
  \hat{w} (\hat{\Delta}^+ ) = \hat{\Delta}^+ \} .
\label{D}
\ee
With this preparation the next lemma holds;
\begin{lem} \label{lem1}
\begin{enumerate}
\item
  $\tilde{W} = D \sdprod \hat{W} $,
 namely, any element $\hat{w}$ of $\tilde{W}$ is uniquely expresseible
as $\hat{w} = \hat{\tau} \hat{w_{0}} $,
$\hat{\tau} \in D$, $\hat{w_0} \in \hat{W}$,
and $\hat{W}$ is a normal subgroup of $\tilde{W}$.
\item
 For $^{\forall} \gamma \in P$, define the element  $\hat{\tau}_{\gamma}$
  of $D$ by the above unique decomposition
  $t_{\gamma} = \hat{\tau}_{\gamma} \hat{w}_{\gamma} $,
  $\hat{\tau}_{\gamma}\in D$,
   $\hat{w}_{\gamma} \in \hat{W}$,
  then it holds that
$$
\ba{l}
\hat{\tau}_{\gamma + \al}
= \hat{\tau}_{\gamma}  ~~~\mbox{for $^{\forall} \al \in Q$},\\
\hat{\tau}_{\gamma_1 + \gamma_2} = \hat{\tau}_{\gamma_1}
\hat{\tau}_{\gamma_2}, \\
\mbox{the map } \gamma \in P ~ \longmapsto ~ \hat{\tau}_{\gamma} \in D~
\mbox{is onto.}
\ea
$$
\end{enumerate}
\end{lem}
This lemma implies that  the group $D$ is isomorphic
to $P/Q (\cong Z(G))$ as abelian group.
It is further known \cite{KW} that
$\mbox{Aut}(\hat{\Pi}) \cong \mbox{Aut}(\Pi) \sdprod D$ holds,
so one may call $D$ as the ``group of proper extended
Dynkin dyagram automorphisms''.

                 One can think $\tilde{W}$ as a subgroup of
$\mbox{Aut}(\hat{\gerg})$.
In fact, the action of $\tilde{W}$ on the CSA $\hat{\gert}$
(and its dual $\hat{\gert^*})$ is already defined.
Introduce a Cartan-Weyl base of $\hat{\gerg}$:
$e_{\al +n \de}=e_{\al}z^n \in \hat{\gerg}_{\al +n \de}$,
$e^{\al +n \de}=e^{\al}z^{-n} \in \hat{\gerg}_{-\al -n \de}$
for $\al +n \de \in \hat{\De}_{+}^{\msc{real}}$
and $hz^n \in \gerg_{n \de}$ for $n \in \bz \setminus \{ 0 \}$.
Here
$e_{\al} \in \gerg_{\al}$, $e^{\al} \in \gerg_{-\al}
(\al \in \De_+)$ are the Cartan-Weyl base of $\gerg$ and
$h \in \gert$.
Then $\hat{w}=wt_{\gamma} \in \hat{W}~ (w \in W, t_{\gamma}\in P)$
will act on them as
\bea
\hat{w}(e_{\al}z^n)&=&e_{w(\al)}z^{n-(\al,\gamma)} \nonumber \\
\hat{w}(e^{\al}z^{-n})&=&e^{w(\al)}z^{-n+(\al,\gamma)} \\
\hat{w}(hz^n)&=&w(h)z^n.\nonumber
\eea
It is easy to check that these definitions besides
(\ref{translation2}), (\ref{cwtg}) give  an automorphism
of $\hat{\gerg}$.

Let us turn our interests to the $\gerg$-current algebra with level
$k\in \bz_{> 0}$ (the integrable representations of $\hat{\gerg}$);
The current $\dsp J(z) = J^{\gert}(z) + \sum_{\al \in \De^+}
\{J^{\al}(z) e_{\al} + J_{\al}(z)e^{\al}\}$
is acting on the Hibert space
$\dsp \ca{H}_k = \sum _{\hat{\La} \in \hat{P}_+^k} L(\hat{\La})$,
where $L(\hat{\La})$ is the integrable $\hat{\gerg}$-module with
the highest weight $\hat{\La} $.
We prepare the next notation;
\be
 \mbox{for } ^{\forall} \bm{x} = u(z) + aK ,~ (u(z) \in L\gerg) ~~~
 J \lb \bm{x} \rb \df \frac{1}{2\pi i} \oint\, (J(z),u(z))\,dz + ak .
\label{current alg}
\ee
The following theorem is important for our purpose;
\begin{thm} \label{thm 1}
For an arbitrary element $\hat{w} = w t_{\gamma} $ of $\tilde{W}$
($w \in W$, $\gamma \in P$),
there exists a unitary transformation $U_{\hat{w}}$ on $\ca{H}_k$
such that
\be
 U_{\hat{w}}\, J \lb \bm{x} \rb \, U_{\hat{w}}^{~-1} =
   J\lb \hat{w} (\bm{x}) \rb ,
\ee
namely,
\bea
   U_{\hat{w}}\, J_{\al}(z) \, U_{\hat{w}}^{~-1}&=&
   z^{-(\al,\gamma)}J_{w(\al)}(z), ~~~(^{\forall} \al \in \De^+) \nonumber \\
   U_{\hat{w}}\, J^{\al}(z) \, U_{\hat{w}}^{~-1}&=&
   z^{(\al,\gamma)}J^{w(\al)}(z), ~~~(^{\forall} \al \in \De^+)\nonumber \\
   U_{\hat{w}}\, ( h, J^{\gert}(z) ) \, U_{\hat{w}}^{-1}&=&
   ( w(h) , J^{\gert}(z) )
   -k \langle \gamma , h\rangle  \frac{1}{z}, ~~~(^{\forall} h \in \gert)
                             \nonumber \\
   U_{\hat{w}} | \hat{\La} , \hat{\La} \rangle
      &=& | \hat{\tau}_{\gamma} (\hat{\La}) , \hat{w}(\hat{\La}) \rangle
\eea
where
$\hat{\tau}_{\gamma}\in D$ is determined by the unique decomposition
$t_{\gamma}= \hat{\tau}_{\gamma} \hat{w}_{\gamma}\in D\sdprod \hat{W}$ in
lemma \ref{lem1}.
\end{thm}
For the Sugawara EM tensor
$\dsp T(z) (\equiv \sum_{n \in \bz}\frac{L_n}{z^{n+2}})
 \df \frac{1}{2(k+g^{\vee})} \nprod (J(z) , J(z) ) \nprod$,
one  can obtain the following transformation formula
by  direct computations.
\begin{cor}\label{tr EM}
Let $\hat{w}$ be as above,
\be
   U_{\hat{w}} \, L_n \, U_{\hat{w}}^{-1}  =  L_n
   - \langle w(\gamma) , J^{\gert}_n \rangle
        + \frac{k}{2} |\gamma|^2 \delta_{n,0}  .
\ee
\end{cor}

Generally $U_{\hat{w}}$ maps $L(\hat{\La})$ to another integrable module
$L(\hat{\tau}_{\gamma}(\hat{\La}))$, and $U_{\hat{w}}$ makes $L(\hat{\La})$
invariant if $\hat{w} \in \hat{W}$, or equivalently, $\gamma \in Q$.
It is worth pointing out that the integrability (or locally nilpotency)
of the representation is crucial for
the existence of $U_{\hat{w}}$ (see \cite{Kac}).
So, in  non-integrable cases
we cannot construct the unitary transformations $U_{\hat{w}}$.

{}~

\subsection{Spectral Flow}

{}~

                Now we would like to introduce the concept of spectral flow
which gives rise to automorphisms of TCA being investigated in
section 3, 4 and has its origin in the Lie algebra automorphisms
given in theorem \ref{thm 1}.

First of all, recall the parabolic decomposition \eqn{parabolic};
\be
\ba{lll}
\gerg &=& \gerh ~ \oplus~ \germ_+~ \oplus~ \germ_- \\
            &=& Z(\gerh)~\oplus
                   ~ \gerh_0 ~ \oplus~ \germ_+~ \oplus~ \germ_- \\
            &=& Z(\gerh)~\oplus~
            (\gerh^{(1)}_0 \oplus \cdots \oplus \gerh^{(r)}_0)
            ~ \oplus~ \germ_+~ \oplus~ \germ_- .
\ea
\label{para}
\ee
According to the above decomposition let us introduce the following
root lattices;
\be
\ba{l}
\dsp Q(\gerh_0) = \bigoplus_{i=1}^{r} Q(\gerh_0^{(i)}),~~~
Q(\gerh_0^{(i)})=\sum_{\alpha \in \Pi_{\gerh_{0}^{(i)}}}\bz \alpha,\\
Q(Z (\gerh))= Q \cap Z(\gerh)^*,
\ea
\ee
where $\Pi_{\gerh_0^{(i)}} (\subset \Pi)$ is the set of simple roots
of $\gerh_{0}^{(i)}$.
The corresponding weight lattices are defined by;
\be
\ba{l}
\dsp P(\gerh_0)= \bigoplus_{i=1}^r P(\gerh_0^{(i)}) ,~~~
P(\gerh_{0}^{(i)})=
\sum_{j=1}^{\msc{rank} \gerh_{0}^{(i)}}
\bz \La_{j}^{(i)} \\
P(Z(\gerh))=
\{\omega \in Z(\gerh)^*~:~(\omega |\beta)
\in \bz~ \mbox{for}~^{\forall}\beta \in Q(Z(\gerh))\},
\ea
\label{wtlas}
\ee
where $\{\La_{j}^{(i)}\}_{j=1}^{\msc{rank} \gerh_{0}^{(i)}}$
is the set of fundamental weights of $\gerh_{0}^{(i)}$.

With respect to the semi-simple part
$\gerh_0 \equiv
\gerh^{(1)}_0 \oplus \cdots \oplus \gerh^{(r)}_0$,
let us denote their affine versions by $\hat{\gerh}_0$,
$\hat{\gerh}_0^{(i)}$, and introduce
the set of roots $\dsp \hat{\De}^+_{\gerh_0} \equiv
\coprod_{i=1}^r \hat{\De}^{(i)+}_{\gerh_0}$.
By using the root and weight lattices
$Q(\gerh_{0}^{(i)})$ and $P(\gerh_{0}^{(i)})$
we can also define
$\hat{W}(\gerh_0) \equiv
\hat{W}(\gerh_0^{(1)})\times \cdots \times
\hat{W}(\gerh_0^{(r)})$,
$\tilde{W}(\gerh_0) \equiv
\tilde{W}(\gerh_0^{(1)})\times
\cdots \times \tilde{W}(\gerh_0^{(r)})$
as
$\hat{W}(\gerh_{0}^{(i)})=W(\gerh_{0}^{(i)})
\sdprod Q(\gerh_{0}^{(i)})$,
$\tilde{W}(\gerh_{0}^{(i)})=W(\gerh_{0}^{(i)})
\sdprod P(\gerh_{0}^{(i)})$ respectively.

Naively we would like to define the spectral flow associated with
the automorphism group $\tilde{W}$.
However, there is a technical problem;
because the symmetry of the $\rho$-sector in (\ref{gauge fixed}) is
described by a $\gerh_0$-current algebra
{\em with negative level,\/} we cannot construct
the automorphism  as that given in theorem \ref{thm 1}.
In order to avoid this difficulty we must  define
the transformations of spectral flow
as the actions of a suitable subgroup of $\tilde{W}$,
which we will denote by $\tilde{D}$,
rather than $\tilde{W}$ itself.
This  subgroup should be defined as;
\be
\tilde{D}=
 \{ \hat{w} = \sigma t_{\gamma} \in \tilde{W} ~;~
    \hat{w} (\hat{\De}^+_{\gerh_0}) \subset
    \hat{\De}^+_{\gerh_0},~ \sigma \in W(\gerh_0), ~
    \gamma \in P ~ \}.
\label{tilde D}
\ee
Namely $\hat{w} \in \tilde{D}$ means that  $\hat{w}$ acts
along the $\hat{\gerh}_0$-direction as a diagram automorphism,
and  also notice that
$\hat{w} (\hat{\De}_{\germ}) \subset \hat{\De}_{\germ}$ holds.
So, by considering the action of  $\tilde{D}$, we will be
able to get a well-defined spectral flow without the difficulty
of negative level.

For our porpose it is more convenient to consider
the following subset $\ca{P}(\gerg/\gerh)$ of $P$
instead of $\tilde{D}$ itself;
\be
\ca{P}(\gerg/\gerh)= \{~ \gamma \in P ~;~ ^{\exists} \sigma \in W(\gerh_0),~
   \mbox{s.t} ~ \sigma t_{\gamma} \in \tilde{D} ~\} .
\label{Pgh}
\ee
It is easy to see that,
for $^{\forall} \gamma \in \ca{P}(\gerg/\gerh)$,
$\sigma$ in the R.H.S of \eqn{Pgh} is unique,
and we will denote it by $\sigma_{\gamma}$.
In other words, the set $\ca{P}(\gerg/\gerh)$
has a one-to-one correspondence with $\tilde{D}$.
We write  this correspondence by
\be
\ca{P}(\gerg/\gerh) \ni \gamma ~ \longmapsto ~
   \hat{w}(\gamma)\df  \sigma_{\gamma} t_{\gamma} \in \tilde{D} .
\label{hat w}
\ee
An equivalent defintion of $\ca{P}(\gerg/\gerh)$
which is more geometrical is given by
\be
\ca{P}(\gerg/\gerh)=\{ ~
\gamma \in P ~;~
^{\exists}\sigma \in W(\gerh_0) ~\mbox{s.t} ~
\sigma (C_{0,\gerh_0}^{\msc{aff}}+\gamma)=C_{0,\gerh_0}^{\msc{aff}}
{}~ \},
\label{Pgh 2}
\ee
where  $C_{0, \gerh_{0}}^{\msc{aff}}$ is the subdomain of
$\gert^{*}$
which contains the Weyl alcove of $\gerh_0$;
\bea
C_{0,\gerh_0}^{\msc{aff}}= \{ ~u \in \gert^* ~;~
(u,\al ) \geq 0 ~ (^{\forall} \al \in \De^+_{\gerh_0}),~
 (u,\, ^{\forall}\theta^{(i)})\leq 1 ~ \}. \nonumber
\eea
Here we denote  the maximal root of $\gerh_{0}^{(i)\bc}$ by
 $\theta^{(i)}$ $(1\leq i \leq r)$.
Notice that $\sigma \in W(\gerh_0)$ in the R.H.S of \eqn{Pgh 2}
is the same element as that of \eqn{Pgh}, i.e. $\sigma = \sigma_{\gamma}$.
For instance, in the case of $G=SU(N)$ (and $H$ is arbitrary),
we can find
\bea
\ca{P}(\gerg/\gerh) =\{~\gamma \in P~;~
(\gamma,\, ^{\forall}\al^{(i)}_{l_i}) = 0 , -1 ,~
\mbox{and } \sum_{l_i=1}^{\msc{rank} \gerh_0}
(\gamma,\al^{(i)}_{l_i})\geq -1
{}~\mbox{for each}~ \gerh_0^{(i)}~ \}, \nonumber
\eea
where $\al^{(i)}_{l_i} \in \Pi_{\gerh_0^{(i)}}$
 $(1 \leq l_i \leq \mbox{rank} \gerh_0^{(i)})$.

To proceed further let us consider
the quotient weight lattice $P/Q(\gerh_0)$.
Take any element $\lb \gamma \rb \in P/Q(\gerh_0)$,
it is easy to show from \eqn{Pgh} (or \eqn{Pgh 2})
that the set
$\lb \gamma \rb \cap \ca{P}(\gerg/\gerh)$ necessarily includes
a unique  non-zero element. We will denote it by $\gamma_0$.
In this way one can obtain a one-to-one correspondence
$\lb \gamma \rb \llerarr \gamma_0$ between
$P/Q(\gerh_0)$ and $\ca{P}(\gerg/\gerh)$.
Making use of this correspondence,
we will define an addition (we express it by ``$\dplus$'')
on $\ca{P}(\gerg/\gerh)$ induced from that of $P/Q(\gerh_0)$.
Notice that, on $P(Z(\gerh))\cap P (\subset \ca{P}(\gerg/\gerh))$,
$\dplus$ coincides with the usual addition $+$, but generally does not.
Under these preparations we can prove;
\bea
\hat{w}(\gamma_1 \dplus \gamma_2)
= \hat{w}(\gamma_1)\hat{w}(\gamma_2). \nonumber
\eea

                      We summarize the above results as a proposition.
\begin{prop} \label{prop 1}
The three abelian groups $\tilde{D}$, $\ca{P}(\gerg/\gerh)$
and $P/Q(\gerh_0)$ are isomorphic with one another.
These isomorphisms are given by
\bea
\gamma \in \ca{P}(\gerg/\gerh) &\longmapsto & \hat{w}(\gamma)
                            \in \tilde{D}  \\
\gamma \in \ca{P}(\gerg/\gerh) &\longmapsto & \lb \gamma\rb
                            \in P/Q(\gerh_0).
\eea
In addition, the subgroups $\tilde{D} \cap \hat{W}$,
$\ca{P}(\gerg/\gerh)$ and $Q/Q(\gerh_0)$ are also isomorphic.
\end{prop}

                            We have arrived at the stage to
present the explicit definition  of  spectral flow.
As was already mentioned,
we shall give it associated with
each element of $\ca{P}(\gerg/\gerh)$
or equivalently $\tilde{D}$.
Recall that
the ingredients of the model in the text are
$g, \rho, X, \chi \psi, \zeta \xi$.
$g,\rho$ and $X$ are the variables of
$G$, $H_0^{\bc}/H_0$ and
$Z(H)^{\bc}/Z(H)$-WZW models.
$\{ \left(\chi_{\alpha}, \psi^{\al}\right)\}_{\al \in \De_{\germ}^+}$
are $G/H$ ghost systems.
$\left\{ (\zeta_{\alpha},\xi^{\al}),
(\zeta^{\alpha}, \xi_{\al})\right\}_{\al \in \De^+_{\gerh_0}}$ and
$(\zeta^{\gert} , \xi^{\gert} )$,
are the $\gerh^{\bc}=\gerh_0^{\bc}\oplus  Z(\gerh)^{\bc}$ ghost systems.
Their mode expansions are given by
$\dsp \chi_{\alpha}(z)=\sum_{n}\chi_{\al,n}z^{-n-1},
\psi^{\al}(z)=\sum_{n}\psi^{\al}_{n}z^{-n}$ etc.

                Fix an arbitrary element $ \gamma $
of $\ca{P}(\gerg/\gerh)$.
The spectral flow $\ca{U}_{\gamma}$ should be defined as follows;
\begin{description}
\item[$g$-sector;]
\be
 \ca{U}_{\gamma} = U_{\hat{w}(\gamma)}~~~ \mbox{in theorem \ref{thm 1}}.
\ee
\item[$\rho$-sector;]
Introduce the similar notations $J_{\rho}^{(i)}\lb \cdots \rb$ as
\eqn{current alg} for the current algebra $J_{\rho}^{(i)}(z)$
(the $H^{(i)}_{0,-(k+g^{\vee}+h^{\vee}_i)}$-current algebra);
\bea
\ba{l}
 \mbox{for } ^{\forall} \bm{x}=u(z) + aK^{(i)} ~
   (u(z) \in L\gerh_0^{(i)}, K^{(i)} ~\mbox{is the central element of }
     \hat{\gerh}_0^{(i)}),   \\
 \dsp J_{\rho}^{(i)} \lb \bm{x} \rb
   \df \frac{1}{2\pi i} \oint\, (J_{\rho}^{(i)}(z),u(z))\,dz
   + a (-k-g^{\vee}-h^{\vee}_i) .
\ea
 \nonumber
\eea
We then define
\be
 \ca{U}_{\gamma}\, J_{\rho}^{(i)} \lb \bm{x} \rb \,
 \ca{U}_{\gamma}^{~-1} =
   J_{\rho}^{(i)}\lb \hat{\tau}_{\gamma}^{\gerh_0^{(i)}} (\bm{x}) \rb ,
\ee
where $\hat{\tau}_{\gamma}^{\gerh_0} \equiv
\hat{\tau}_{\gamma}^{\gerh_0^{(1)}} \cdots
\hat{\tau}_{\gamma}^{\gerh_0^{(r)}}$ is the element of $D(\gerh_0)$
given by
\bea
\hat{\tau}^{\gerh_0}_{\gamma}
= \hat{w}(\gamma)|_{\hat{\gerh}_0}, \nonumber
\eea
or  more explicitely,
\be
\ba{rll}
   \ca{U}_{\gamma}\, J^{(i)}_{\rho, \al}(z) \, \ca{U}_{\gamma}^{~-1}&=&
   z^{-(\al,\gamma)} J^{(i)}_{\rho, \sigma_{\gamma}(\al)},
               ~~~(^{\forall} \al \in \De^{(i)+}_{\gerh_0}) \nonumber \\
   \ca{U}_{\gamma}\, J^{(i)\al}_{\rho}(z) \, \ca{U}_{\gamma}^{~-1}&=&
   z^{(\al,\gamma)}J^{(i)\sigma_{\gamma}(\al)}_{\rho}(z),
              ~~~(^{\forall} \al \in \De^{(i)+}_{\gerh_0}) \nonumber \\
   \ca{U}_{\gamma}\, ( h, J^{(i)\gert}_{\rho}(z))  \,
   \ca{U}_{\gamma}^{~-1}&=& \dsp
   ( \sigma_{\gamma}(h) , J^{(i)\gert}_{\rho}(z) )
   +(k+g^{\vee}+h^{\vee}_i) \langle \gamma^{(i)}, h \rangle \frac{1}{z},
    ~~~(^{\forall} h \in \gert(\gerh_0^{(i)})) . \\
    \ca{U}_{\gamma} |\hat{\la},\hat{\la} \rangle_{\rho} & = &
       | \hat{\tau}_{\gamma}^{\gerh_0} (\hat{\la}),
          \hat{\tau}_{\gamma}^{\gerh_0} (\hat{\la}) \rangle_{\rho}
\ea
\label{sflow jrho}
\ee
\item[$X$-sector;]
\be
\ba{l}
  \ca{U}_{\gamma} \, (h, J_{X}(z)) \, \ca{U}_{\gamma}^{~-1}
  \dsp  = (h,J_{X}(z)) + (k+g^{\vee}) \langle \gamma,h \rangle \frac{1}{z},
    ~~~ (^{\forall} h \in Z(\gerh))\\
  \ca{U}_{\gamma} | \beta\rangle_X =
          |\beta - (k+g^{\vee})\gamma \rangle_X ,
\ea
\label{sflow jX}
\ee
  where the primary state $|\beta\rangle_X $ $(\beta \in Z(\gerh)^*)$
  is defined by
\bea
   (h, J_{X,n})| \beta \rangle_X = \delta_{n,0} \langle \beta, h \rangle
                                                       |\beta\rangle_X,
   ~~~ (n \in \bz_{\geq 0}, ~^{\forall} h \in Z(\gerh)). \nonumber
\eea
\item[$\chi\psi$-sector;]
 Define the following sets of affine roots;
\bea
 \hat{\Delta}_{\germ}=
 \{ \alpha +n \delta~;~ \alpha \in \Delta_{\germ}^{+} \coprod
 \Delta_{\germ}^{-} ,~ n \in \bz \},  ~
 \hat{\De}_{\germ}^{\pm}
   = \hat{\De}_{\germ}\cap \hat{\De}^{\pm}, \nonumber
\eea
and introduce the following notation for the mode expansions
of $\chi\psi$;
\bea
     \bm{\psi}_{\hat{\al}} =
     \left \{ \ba{ll}
             \psi^{-\al}_n   ~~ & \al \in \De_{\germ}^- \\
             \chi_{\al,n} ~~ & \al \in \De_{\germ}^+
    ~~~  (^{\forall} \hat{\al} \equiv \al + n\de \in \hat{\De}_{\germ})
   \ea \right. . \nonumber
\eea
We then define
\be
  \ba{l}
     \ca{U}_{\gamma}\, \bm{\psi}_{\hat{\al}} \,
     \ca{U}_{\gamma}^{~-1}
             = \bm{\psi}_{\hat{w}(\gamma)(\hat{\al})},
              ~~~(^{\forall} \hat{\al} \in \hat{\De}_{\germ}) \\
      \dsp   \ca{U}_{\gamma} |0 \rangle_{\chi\psi} =
           \prod_{\hat{\al} \in \Phi_{\hat{w}(\gamma)}}
      \bm{\psi}_{- \hat{\al}} |0 \rangle_{\chi\psi} ,
  \ea
\ee
   where $\Phi_{\hat{w}(\gamma)}$ is the subset of $\hat{\Delta}^+$
   given by
   \bea
     \Phi_{\hat{w}}=
          \hat{w}(\hat{\Delta}^{-}) \cap \hat{\Delta}^{+},~~
            \mbox{for~}^{\forall} \hat{w} \in \tilde{W}. \nonumber
   \eea
The Fock vacuum $|0\rangle_{\chi \psi}$ is the state which satisfies
\bea
\bm{\psi}_{\hat{\al}}|0\rangle_{\chi \psi}=0, ~~~
(^{\forall} \hat{\al} \in \hat{\De}_{\germ}^{+}) . \nonumber
\eea
\item[$\zeta\xi$-sector;]
Introduce the following notations;
\bea
\ba{l}
   \bm{\zeta}_{\al +n \de}
   = \left\{ \ba{ll}
           \zeta_{\al,n} ~~ & \al \in \De^+_{\gerh_{0}} , \\
           \zeta^{-\al}_{n} ~~ & \al \in \De^-_{\gerh_{0}} ,\\
        \ea \right.   \\
   \bm{\xi}_{\al +n \de}
   = \left\{ \ba{ll}
           \xi_{\al,n} ~~ & \al \in \De^+_{\gerh_0} , \\
           \xi^{-\al}_{n} ~~ & \al \in \De^-_{\gerh_0} . \\
        \ea \right.  ,
\ea
\nonumber
\eea
and then we define
\be
\ba{l}
\ca{U}_{\gamma} \, \bm{\zeta}_{\hat{\al}} \,
\ca{U}_{\gamma}^{~-1}
   = \bm{\zeta}_{\hat{\tau}_{\gamma}^{\gerh_0}(\hat{\al}) }    ,\\
\ca{U}_{\gamma} \, \bm{\xi}_{\hat{\al}} \,
\ca{U}_{\gamma}^{~-1}
   = \bm{\xi}_{\hat{\tau}_{\gamma}^{\gerh_0}(\hat{\al}) }    ,
{}~~~(^{\forall} \hat{\al} \in \hat{\De}_{\gerh_0}) .
\ea
\label{sflow zeta xi 1}
\ee
For the Cartan directions we set the actions of $\ca{U}_{\gamma}$
as
\be
\ba{l}
\ca{U}_{\gamma} \, (h, \zeta^{\gert}_n) \, \ca{U}_{\gamma}^{~-1}
   = (\sigma_{\gamma}(h) , \zeta^{\gert}_n) ,~~~ \\
\ca{U}_{\gamma} \, (h ,\xi_n^{\gert}) \, \ca{U}_{\gamma}^{~-1}
   = (\sigma_{\gamma}(h)  , \xi_n^{\gert}), ~~~(^{\forall} h \in \gert).
\ea
\label{sflow zeta xi 2}
\ee
The Fock vacuum $|0\rangle_{\zeta \xi}$ is the state
which satisfies
\bea
\ba{l}
\bm{\zeta}_{\al +n \de}|0\rangle_{\zeta \xi}=
  (h,\zeta^{\gert}_n)|0\rangle_{\zeta\xi} =0 ~~~
    (^{\forall} n\geq 0,~ ^{\forall} \al \in \De_{\gerh_0},
   ~^{\forall} h \in \gert), \\
\bm{\xi}_{\al +n \de}|0\rangle_{\zeta \xi}=
  (h,\xi^{\gert}_n)|0\rangle_{\zeta\xi} =0 ~~~
    (^{\forall} n> 0,~ ^{\forall} \al \in \De_{\gerh_0},
  ~ ^{\forall} h \in \gert) ,
\ea
\nonumber
\eea
and its transformation law which is consistent with \eqn{sflow zeta xi 1},
\eqn{sflow zeta xi 2} is as follows;
\be
\ca{U}_{\gamma} | 0 \rangle_{\zeta\xi}=
\prod_{\al \in \De^+_{\gerh_0}}
\bm{\zeta}_{-\hat{\tau}_{\gamma}^{\gerh_0}(\al)} \,
\bm{\xi}_{\al} |0 \rangle_{\zeta\xi}.
\ee
\end{description}

             For the coset part ($g$, $\chi\psi$) these definitions coincide
with those given in \cite{LVW,HT}.
But one must also introduce the concept
of spectral flow for the $H_0^{\bc}$ and $Z(H^{\bc})$-parts
in order to formulate the model {\em as a Lagrangian field theory.}

                 Our main results in this appendix
are included  in  the following theorem;
\begin{thm} \label{main thm}
\begin{enumerate}
\item
All the BRST charges ($Q_{G/H}$, $Q_{Z(H^{\bc})}$, $Q_{H_0^{\bc}}$,
$Q_{\msc{tot}}$) which are defined in section 2 are invariant under
the spectral flows, that is, for an arbitrary element $\gamma$ of
$\ca{P}(\gerg/\gerh)$,
\be
  \ca{U}_{\gamma}\, Q_* \, \ca{U}_{\gamma}^{~-1} = Q_* ,
\ee
where ``$*$'' means ``$G/H$'' or ... ``tot''.
\item
The total EM tensor \eqn{totalT} is invariant under the spectral flow
modulo BRST-exact terms,
explicitely written as;
\bea
\ca{U}_{\gamma} \, L_{\msc{tot},n} \, \ca{U}_{\gamma}^{~-1}
 & = & L_{\msc{tot},n} - \langle \sigma_{\gamma}(\gamma) ,
      J^{\gert}_{\msc{tot},n} \rangle \nonumber \\
  & = & L_{\msc{tot},n} - \{ Q_{\msc{tot}}, \sqrt{2} \al_+
      \langle \sigma_{\gamma}(\gamma)  , \zeta_n \rangle \} .
\eea
\item
$\{ G^{\pm}_{G/H},T_{G/H},J_{G/H} \}$,
the TCA of Kazama-Suzuki sector given in section 2,
is `strictly'' invariant
under the spectral flow;
\be
\ba{l}
\ca{U}_{\gamma}\, T_{G/H}(z)\,\ca{U}_{\gamma}^{~-1}
= T_{G/H}(z) ,~~~
\ca{U}_{\gamma}\, G^{\pm}_{G/H}(z)\,\ca{U}_{\gamma}^{~-1}
= G^{\pm}_{G/H}(z) ,  \\
\ca{U}_{\gamma}\, J_{G/H}(z)\,\ca{U}_{\gamma}^{~-1}
= J_{G/H}(z) .
\ea
\ee
\end{enumerate}
\end{thm}
This theorem implies that the BRST-cohomology states of our model
are invariant under the spectral flow.
This fact plays a crucial role in our discussions in section 4.
Any physical state $|\Psi \rangle$
and its transformed state $\ca{U}_{\gamma}|\Psi \rangle$ possess
the same KS $U(1)$-charges.

                   We will summarize
a few useful  statements as propositions.
\begin{prop} \label{prop 2}
The map $ \gamma  \in \ca{P}(\gerg/\gerh) \longmapsto
\ca{U}_{\gamma}$ gives a homomorphism of abelian group.
Namely,
\be
\ca{U}_{\gamma_1  \scdplus \gamma_2}
  =  \ca{U}_{\gamma_1} \ca{U}_{\gamma_2} ,~~~
    (^{\forall} \gamma_1, \gamma_2 \in \ca{P}(\gerg/\gerh)) .
\ee
\end{prop}

           Let us introduce the following subset of $\hat{W}$;
\be
\hat{W}(\gerg/\gerh)=
\{ \hat{w} \in \hat{W} ~;~ \Phi_{\hat{w}} \subset \hat{\Delta}^+_{\germ} \},
\ee
and then the next propositions hold;
\begin{prop} \label{prop 3}
If $\hat{w} \in \hat{W}(\gerg/\gerh)$,
$\gamma \in \ca{P}(\gerg/\gerh)$, then
$\hat{w}(\gamma) \hat{w} \hat{\tau}_{\gamma}^{-1} \in \hat{W}(\gerg/\gerh)$
holds.
Especially, for any $\al \in \ca{P}(\gerg/\gerh) \cap Q$,
$\hat{w}(\al) \hat{w} \in \hat{W}(\gerg/\gerh)$.
\end{prop}
\begin{prop} \label{prop 4}
\be
   \hat{W}(\gerg/\gerh) = (\tilde{D}\cap \hat{W}) \, W(\gerg/\gerh) ,
\ee
where we set
\bea
W(\gerg/\gerh)=
  \{ w \in W ~;~ \Phi_{w} \subset \Delta^+_{\germ} \} .  \nonumber
\eea
\end{prop}
(Refer \cite{HT} for the proof)
This means that  any element $\hat{w}$ of $\hat{W}(\gerg/\gerh)$
can be uniquely decomposed  as $\hat{w} = \om w$,
where $\om \in \tilde{D}\cap \hat{W}$ and
$w \in W(\gerg/\gerh)$,
or equivalently
one can uniquely express it as $\hat{w} = \hat{w}(\al) w $
by an element   $\al \in \ca{P}(\gerg/\gerh)\cap Q$.

\end{document}